\documentclass{aa}
\usepackage[varg]{txfonts}
\usepackage{amsmath}
\usepackage{graphicx}
\usepackage{url}
\usepackage{listings}
\usepackage{xcolor}
\usepackage{tabu}
\usepackage{units}
\usepackage{wasysym}
\usepackage[font={small},labelfont=bf,belowskip=2mm,aboveskip=1mm]{caption}
\usepackage{geometry}
\usepackage{multirow}
\usepackage{leftidx}
\usepackage[hidelinks]{hyperref}
\usepackage{csquotes}
\bibpunct{(}{)}{;}{a}{}{,} % to follow the A&A style

\def\ab{a_i}
\def\abmax{\ab^\text{max}}
\def\abmin{\ab^\text{min}}

\def\abtrend{\ab^{\text{trend}} \left( T_0 \right)}
\def\aoffset{a^\mathrm{offset}}

\def\abint{\ab^{\text{int}}}
\def\erand{e_i}
\def\esyst{e_i^\text{syst}}
\def\si{\sigma_i}

\def\sint{\sigma_{\text{X, int}}}
\def\sintfe{\sigma_{\text{Fe, int}}}
\def\sintti{\sigma_{\text{Ti, int}}}
\def\sintmg{\sigma_{\text{Mg, int}}}
\def\nstar{n_{\text{star}}}

\def\nh{N_{\text{H}}}
\def\nx{N_{\text{x}}}

\def\teff{T_\text{eff}}

\def\logg{\log{g}}
\def\feh{\left[ \mathrm{Fe} / \mathrm{H} \right]}
\def\mh{\left[ \mathrm{M} / \mathrm{H} \right]}
\def\afe{\left[ \alpha / \mathrm{Fe} \right]}
\def\vmic{v_{\mathrm{mic}}}
\def\vmicdwarf{v_{\mathrm{mic, dwarf}}}
\def\vmicgiant{v_{\mathrm{mic, giant}}}
\def\vmac{v_{\mathrm{mac}}}

\def\snr{\mathrm{S} / \mathrm{N}}

\def\FPR{\mathrm{FPR}}
\def\FPRmax{\FPR_\mathrm{max}}

\def\aicc{\mathrm{AIC_c}}
\def\daicc{\Delta \aicc}
\def\padd{P \left(H_\mathrm{Add\-Mix} \right)}

\title{Atomic diffusion and mixing in old stars\\VII. Abundances of Mg, Ti, and Fe in M30\thanks{Based on data collected at the ESO telescopes under programme 099.D-0748(A)}}
\titlerunning{Atomic diffusion and mixing in old stars VII}
\author{Alvin~Gavel\inst{\ref{inst:uppsala}} \and Pieter Gruyters\inst{\ref{inst:uppsala}} \and Ulrike Heiter\inst{\ref{inst:uppsala}} \and Andreas J.~Korn\inst{\ref{inst:uppsala}} \and Thomas Nordlander\inst{\ref{inst:anu},\ref{inst:astro3d}} \and Kilian H. Scheutwinkel\inst{\ref{inst:uppsala}} \and Olivier A. Richard\inst{\ref{inst:montpellier}}}
\institute{Observational Astrophysics, Division of Astronomy and Space Physics, Department of Physics and Astronomy, Uppsala University, Box 516, 75120 Uppsala, Sweden\label{inst:uppsala} \and Research School of Astronomy and Astrophysics, Australian National University, Canberra, ACT 2611, Australia\label{inst:anu} \and ARC Centre of Excellence for All Sky Astrophysics in 3 Dimensions (ASTRO 3D)\label{inst:astro3d} \and Laboratoire Univers et Particules de Montpellier, Université de Montpellier, CNRS, Place Eugène Bataillon, 34095 Montpellier, France \label{inst:montpellier}}
\date{Received 9 March 2021 / accepted 24 May 2021}

\abstract
{}
{We attempt to constrain the efficiency of additional transport or mixing processes that reduce the effect of atomic diffusion in stellar atmospheres.}
{We apply spectral synthesis methods to spectra observed with the GIRAFFE spectrograph on the VLT to estimate abundances of Mg, Ti, Fe, and Ba in stars in the metal-poor globular cluster M30. To the abundances we fit trends of abundances predicted by stellar evolution models assuming different efficiencies of additional transport or mixing processes. The fitting procedure attempts to take into account the effects of parameter-dependent systematic errors in the derived abundances.}
{We find that the parameter $T_0$, which describes the efficiency of additional transport or mixing processes, can almost certainly be constrained to the narrow range $\log_{10}{\left( T_0 / \left[ \unit{K} \right] \right)}$ between $6.09$ and $6.2$. This corresponds to decreased abundances for stars at the main sequence turn-off point compared to the red giant branch by $0.2\,\unit{dex}$ for Mg, $0.1\,\unit{dex}$ for Fe, and $0.07\,\unit{dex}$ for Ti. We also find that while our estimates do have non-negligible systematic errors stemming from the continuum placement and the assumed microturbulence, our method can take them into account.}
{Our results partly amend the results of an earlier paper in this article series, that tentatively used a value of $\log_{10}{\left( T_0 / \left[ \unit{K} \right] \right)} = 6.0$ when modelling the Spite plateau of lithium. To more easily distinguish physical effects from systematic errors, we recommend that studies of this kind focus on elements for which the expected surface abundances as functions of effective temperature have a distinct structure and cover a wide range.}

\begin{document}

\maketitle

\section{Introduction}
% First approximation: abundances are constant
To first approximation, one might expect the atmospheres of the stars in a stellar cluster to have identical elemental abundances: They formed from the collapse of a chemically almost homogeneous gas cloud, and the outer layers of a star are mostly unaffected by nucleosynthesis. Closer inspection has shown that this is not quite true: The assumptions that cluster stars form from chemically identical gas and that nucleosynthesis has no effect on observed abundances, are both violated in some circumstances. There are also several processes that cause elements to separate by depth in the stellar atmosphere. This paper focuses on constraining a parameter that describes one of those processes.

% Not quite true because 
The assumption that the stars in a cluster are formed from chemically homogeneous gas would be true if they all formed at exactly the same time. In reality, the formation of a stellar cluster takes time, and by the time the last stars form, several of the first stars have already reached the end of their evolution and have polluted the remaining gas. This causes the phenomenon of anticorrelations in stellar clusters, where increased abundances of certain elements are statistically correlated with lowered abundances of others, such as Mg with Al and Na with O~\citep{anticorrelations_ona_mgal}. Such anticorrelations were first described in \citet{anticorrelation_f1rst}, and in \citet{anticorrelations_carretta} they were definitively shown to be correlated with the orbital parameters of the stars. This led to the hypothesis that the anticorrelations somehow reflect chemical conditions at the site of formation, which in turn are probably mostly affected by material ejected from the first generation of stars.

The assumption that nucleosynthesis does not affect surface abundances is violated in giant branch stars through the process of dredge-up, in which the convective envelope reaches deep enough into the star to start pulling up materials produced by core hydrogen burning. However, this mainly affects the isotopic ratios of C, N, and O and the abundances of Li and Be~\citep{pagel}, none of which we measure in the main part of this study.%, although we replicate a study of Li from a previous paper in Appendix~\ref{app_paper_6}.

% Second approximation: diffusion
Several different processes cause elements to separate with depth in the stellar atmosphere. Some of these processes are well understood and some are still not. The conceptually simplest are the diffusive processes. Over long timescales, there are two dominant processes that tend to cause different species to drift vertically in a macroscopically stationary region of a stellar atmosphere: There is the tendency for gravity to pull heavier species downwards, referred to as `gravitational settling'. On the other hand, there is the tendency for radiation to push more opaque species upwards, referred to as `radiative acceleration'. In models of old metal-poor stars, gravitational settling tends to be strongest, creating an overall tendency to lower the abundances of metals in the outer layers of the atmosphere, except for some elements that undergo significant levitation around the turn-off point~(TOP). 
%Counteracting all of this is the tendency for random brownian motion to restore a uniform distribution, referred to as \emph{thermal diffusion}. 
The separation of elements is counteracted by macroscopic vertical mass flows -- that is, convection~\citep{Michaud70,Richard01}. Based on observations, it is known that warm stars have noticeably lower abundances than giant stars. Based on modelling, this is thought to be because warm stars have smaller convective envelopes, while giant stars have large convective envelopes that keep their abundances close to the original values. An overview of this topic is given in the first paper in this article series,~\citet{same_1}.

% Third approximation: Add\-Mix
%% Describe observations
However, there appears to be at least one more mechanism that affects the separation of elements in the atmosphere. If only the diffusive processes described above, together with convection were important, we would expect to see much stronger trends in abundances as a function of effective temperature than have actually been observed in stellar clusters~\citep{same_1}. This implies that some other transport process of chemical species is counteracting the effects of diffusion~\citep{Michaud76, Richard01}. The most conspicuous case is lithium, for which there is a wide range of $\teff$, called the `Spite plateau', for which abundances are almost constant~\citep{Spite, Richard_metalpoor_1}.
%% Aside about naming
In this series of papers, we refer to this mechanism as the `additional transport or mixing process'~(Add\-Mix). Most other papers use the term `turbulent mixing' instead. Add\-Mix should not be confused with the term `extra mixing', which is sometimes used for the unrelated process of thermohaline mixing~\citep{thermohaline}.
%% Theoretical understanding of Add\-Mix
As the vague name is meant to emphasise, the physical mechanism of Add\-Mix is currently not well understood~\citep{Richer00}. It has been suggested that it might involve mass loss and rotationally induced mixing~\citep{Michaud83b, Michaud91b, vick13}. Essentially, stellar evolution models incorporating Add\-Mix postulate the existence of some sort of turbulent mixing process parametrised in terms of a turbulent diffusion coefficient $D_T$, defined as
\begin{align}
D_T &\equiv \omega D_{\mathrm{He}} \left( T_0 \right) \left( \frac{\rho \left( T \right)}{\rho \left( T_0 \right)} \right)^{-n}, \label{eq_DT}
\end{align}
where the proportionality constant $\omega$ is a real number, the exponent $n$ is some integer, $\rho$ is density at a given temperature, $T_0$ is a reference temperature and $D_{\mathrm{He}} \left( T_0 \right)$ is the atomic diffusion coefficient of He at the temperature $T_0$~\citep{Richer00, Richard05}. We note that while $D_{\mathrm{He}}$ is included in~\eqref{eq_DT}, based on the physical assumptions made by the modellers, we could be agnostic about what Add\-Mix actually is and treat $\omega D_{\mathrm{He}} \left( T_0 \right)$ as simply being a generic coefficient with dimensions of area divided by time.

% How do we determine the parameters?
Since we lack a fundamental physical model to derive them from first principles, the parameters in $D_T$ have to be determined empirically by fitting model predictions to observations. This has been done by several different studies, including the article series that this paper is part of. Fitting to observed abundances of Li, Be, and B in the Sun has led to the adoption of $n = 3$. This simultaneously leads to adopting the proportionality $\omega = 400$~\citep{Michaud91b, Richard_metalpoor_1, Richard05}. The parameter $T_0$ still needs to be further constrained. By convention, we label Add\-Mix models in terms of $\log_{10}{\left( T_0 / \left[ \unit{K} \right] \right)}$. This is usually done in a shorthand such that T$5.9$ should be read as `the model for which $\log_{10}{ \left( T_0 / \left[ \unit{K} \right] \right)} = 5.9$'. The third paper in this article series -- \citet{same_3}, henceforth called `Paper~III' -- arrived at T$6.0$ for the metal-poor cluster NGC~6397. The sixth paper in this article series -- \citet{same_6}, henceforth called `Paper~VI' -- adopted this value for M30 (NGC-7099), based on their similar metallicities, but noted that T$5.8$ provided a better fit to their derived abundances.

% What is our goal?
The main goal of this article is to provide a more accurate estimate of $T_0$ by comparing abundance trends predicted for different values of $T_0$ to derived abundances for M30. To do this we must also address a problem with interpreting derived abundances: All abundance estimates are affected by systematic errors, which depend to some extent on stellar parameters such as $\teff$. This means that any measured trend will be a difficult-to-disentangle mixture of systematics and the actual physical trend. We have occasionally even encountered scepticism as to what extent measured abundance trends even measure anything other than the systematics of the measurement method. Hence, secondary objectives of the article are to develop a well-defined method for constraining systematic uncertainties, to then use this to definitively show that diffusion processes do have measurable effects, and to clarify how well $T_0$ can be constrained when systematics are taken into account.

\section{Observations}
We have photometric and spectroscopic observations of 177 stars in the globular cluster M30. The cluster has been selected on the basis that it is very metal-poor, with $\mh = -2.35$, and relatively little studied. Since it is $6\,\unit{kpc}$ below the Galactic plane~\citep{M30_where}, it has the fairly low reddening of $E \left( B - V \right) = 0.03 \pm 0.01\,\unit{mag}$~\citep{M30_photom, M30_photom_err}. Together with the clusters NGC~6397, NGC~6752 and M4, which have previously been studied in this article series, it forms a sequence of $\mh$ from $-2.35$ to $-1$, which allows us to probe AddMix as a function of overall metallicity. M30 also has the advantage of being far above the galactic plane, minimising the effect of Galactic reddening.

\subsection{Photometric observations}
%The stars were originally selected based on Strömgren photometry obtained with the Danish 1.54-metre telescope on La Silla~\citep{grundahl_99}. However, no standard stars were observed as part of these observations, which means that they cannot be calibrated well enough to be useful in determining stellar parameters.
The stars had previously been observed with $VI$ broadband photometry by Peter Stetson~\citep{stetson_2000, stetson_2005}. This was used to estimate the stellar parameters needed for abundance determination, as described in Sect.~\ref{sec_SP}. A colour-magnitude diagram can be found in Paper~VI~\citep[Fig. 1]{same_6}.

\subsection{Spectroscopic observations}
The Fibre Large Array Multi Element Spectrograph (FLAMES) is a spectrograph connected to the second Unit Telescope (UT2) at the Very Large Telescope (VLT) facility~\citep{flames}. This feeds the GIRAFFE\footnote{Despite customarily being spelled in uppercase, the name is not an acronym.} spectrograph. The spectra have a resolution $\lambda / \Delta \lambda = 23\,000$ % I guess? That's what the fits header says, while the GIRAFFE webpage cites either 35150 or 24300, depending on whether it's IFU/Argus or MEDUSA, respectively.
and were taken with an exposure time around $2775\,\unit{s}$. They were observed in the HR6 mode, which covers the wavelength range $4538$-$4760\,\unit{\AA}$. They were observed as part of project 099.D-0748(A) with principal investigator Pieter Gruyters. Several spectra were taken of each star, typically around 30 for faint stars and around 5 for bright stars. %We divide the stars into the categories Turn-Off Point (TOP), Sub-Giant Branch (SGB) and Red Giant Branch (RGB). 
We give a full list of the stars in Table~\ref{table_individual_SP}. %Maybe we'll end up putting them in that database instead

For each star, the observed spectra were coadded into a single spectrum. As described in Appendix~\ref{app_spectra}, this was done before correcting for radial velocity, and also required the removal of sky flux. After this the spectra were visually inspected, and those with issues likely to prevent a meaningful abundance determination were flagged and removed from further analysis. For the abundance derivation we further co-added the spectra for individual stars into group spectra of stars with similar stellar parameters, as described in Appendix~\ref{app_combined}.

\section{Stellar evolution models}\label{sec_trend}
\begin{figure}
\centering
\includegraphics[width=8.8cm]{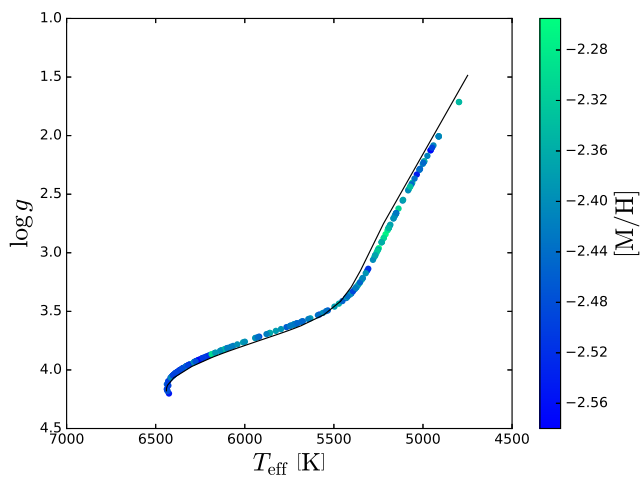}
\caption{Isochrone used in the T$6.09$ model, together with the $\teff$, $\logg$, and $\mh$ of the individual stars estimated as described in Sect.~\ref{sec_SP}.}
\label{ill_kiel}
\end{figure}

We computed stellar evolution models, including atomic diffusion (with radiative accelerations) and Add\-Mix as described by Eq.~\eqref{eq_DT}, using the Montr{\'e}al/Montpellier stellar evolution code~\citep{Turcotte98, Richer00, Richard01}. These were used to compute isochrones describing the evolutionary variation in surface chemical composition, giving us expected abundances as functions of effective temperature calculated for six discrete $T_0$ values: T$5.8$, T$5.9$, T$5.95$, T$6.0$, T$6.09$, and T$6.2$, assuming a cluster age of $13.5~\unit{Gy}$. With this range, we sample the range of diffusion models used in previous papers of this series, as well as those that are compatible with the empirical constraints arising from a thin and flat Spite plateau -- notwithstanding the slight rise in lithium in subgiants prior to the onset of the first dredge-up as observed in NGC~6397,~\citep{lithium_why, same_1, lind_ngc6397}. We show the isochrone in Table~\ref{ill_kiel} together with the estimated stellar parameters of the individual stars, which have been estimated as described in Sect.~\ref{sec_SP}. Previous tests have shown that the isochrones are able to reproduce the colour-magnitude diagram of M92 (NGC 6341), which has similar metallicity to M30~\citep{Vandenberg_2002}.

In these models, the vertical stratification of chemical composition is followed for the elements that have known radiative cross-sections, and evolutionary variations are thus predicted for those elements~\citep{richer_98}. This includes Fe, Ti, and Mg, which our spectra contain lines from. Radiative cross-sections, and hence model predictions, are not available for Ba, which prevented us from fitting to our abundance estimates of that element. We discuss this element in Appendix~\ref{app_unused}.

We show the predicted trends, together with our derived abundances (see Sect.~\ref{sec_abundance_derivation}), for Mg in Fig.~\ref{ill_Mg4702_Full_model_fit}, for Ti in Fig.~\ref{ill_Ti4563_Full_model_fit}, and for Fe in Fig.~\ref{ill_Fe4583_Full_model_fit}. For Fe and Mg, the expectation is that the abundance should increase with decreasing effective temperature, while the expectation for Ti is that it should drop and then rise, with a sharper rise the lower $T_0$. This would seem to make the Ti trend the easiest to distinguish, but the trend is also much weaker than the trends expected for the other elements.

%\begin{figure}
%\centering
%\includegraphics[width=8.8cm]{ill_Fe_delta_ab.eps}
%\caption{Model trends at $13.5~\unit{Gy}$ for iron abundance, under different $T_0$. $\Delta \log \varepsilon \left( \mathrm{Fe} \right)$ denotes the difference in abundance from that of the primordial gas cloud.}
%\label{ill_Fe_ab}
%\end{figure}
%\begin{figure}
%\centering
%\includegraphics[width=8.8cm]{ill_Ti_delta_ab.eps}
%\caption{Model trends at $13.5~\unit{Gy}$ for titanium, under different $T_0$. $\Delta \log \varepsilon \left( \mathrm{Ti} \right)$ denotes the difference in abundance from that of the primordial gas cloud.}
%\label{ill_Ti_ab}
%\end{figure}
%\begin{figure}
%\centering
%\includegraphics[width=8.8cm]{ill_Mg_delta_ab.eps}
%\caption{Model trends at $13.5~\unit{Gy}$ for magnesium, under different $T_0$. $\Delta \log \varepsilon \left( \mathrm{Mg} \right)$ denotes the difference in abundance from that of the primordial gas cloud.}
%\label{ill_Mg_ab}
%\end{figure}

\section{Spectroscopic derivation of abundances}\label{sec_abundance_derivation}
To derive abundances for individual spectra, we use a pipeline based on the spectral synthesis code Spectroscopy Made Easy~(SME), described in Sect.~\ref{sec_sme}. The pipeline is similar to the pipeline for determining stellar parameters described in~\citet{me_1}.

We present abundances using what is informally known as the `12-scale'. That is, we show
\begin{align}
\log \varepsilon \left( \text{x} \right) &\equiv \log_{10}{\left( \nx / \nh \right)} + 12,
\end{align}
where $x$ is some element, $\nx$ is the number density of the element, and $\nh$ is the number density of hydrogen.

\subsection{Algorithm}\label{sec_algorithm}
The algorithm works by fitting a synthetic spectrum to the observed spectrum. This is done in a wavelength segment around the line in question that has been chosen to be wide enough to allow a reliable continuum determination. The fitting is done over a wavelength segment covering the line. %, but avoiding the centre if it is estimated to be deep enough to be affected by deviations from the approximation of Local Thermodynamic Equilibrium, described in Sect.~\ref{sec_nlte}.

The fitting involves three steps. The first step is cleaning of contaminants: Before looking at the observed spectrum, we generate two model spectra: One that contains spectral lines for all elements, and one that contains only spectral lines for the element that we intend to fit. Pixels for which the estimated fluxes differ by more than 0.005 are flagged as contaminated, and do not get used in the fitting.
% Thomas asked if this is relevant at this metallicity. The answer is sometimes. In reproducing the results of paper VI, it turns out that some of the faint lines can't be used at all, since they're too blended.
The second step is normalisation: The continuum level of the observed spectrum is estimated, as described in Sect.~\ref{sec_sme}. The final step is fitting: The model spectrum is fitted to the observed spectrum by minimising a $\chi^2$-like goodness-of-fit metric, as described in Sect.~\ref{sec_sme}. Both the abundance of the element in question, and a Gaussian broadening parameter, are free parameters in the fit. The broadening parameter is mostly intended to capture the effects of macroturbulence at the depth of line formation, but can also capture other physical and instrumental effects.

\subsection{Line selection}
The wavelength region covered by our spectra is relatively line-poor. While this has the benefit of decreasing the number of contaminants, it gave us relatively few lines to work with. We selected one line of Fe, one line of Mg, two lines of Ti, and one line of Ba, shown in~Table~\ref{table_lines_sme} together with line masks as described in Sect.~\ref{sec_sme} and $\log{gf}$ values. The lines are taken from the Gaia-ESO line list described in~\citet{linelist}, which in turn uses values from~\citet{NIST10}, \citet{PTP}, \citet{WLSC}, \citet{K13} and~\citet{MW}. As a shorthand, we will refer to the lines as Mg4702, Ti4563, Ti4571, Fe4583, and Ba4554, respectively. In the main body of this article we only study Mg4702, Ti4563, and Fe4583, but we discuss Ti4571 and Ba4554 in Appendix~\ref{app_unused}.

We note that the Fe~II line at $4583.8292\,\unit{\AA}$ overlaps with a weaker Fe~I line at $4583.7195\,\unit{\AA}$. The blend was taken into account in our main analysis, but was not treated completely consistently in the 3D~analysis described in Appendix~\ref{app_3D_comp}.

\begin{table*}
\caption{Lines used to derived stellar abundances. Only Mg4702, Ti4563 and Fe4583 were used for fitting the trend models. All wavelengths are given in $\unit{\AA}$. The line mask describes the wavelength interval used in the fitting of abundances, and the segment mask describes the entire wavelength region modelled around a line, as described in Sect.~\ref{sec_sme}.}\label{table_lines_sme}
\centering
\begin{tabular}{ c | c | c | c | c | c | c }
\multirow{2}{*}{Species} & Line & \multicolumn{2}{c|}{Line mask $[\unit{\AA}]$}   & \multicolumn{2}{c|}{Segment mask $[\unit{\AA}]$} & \multirow{2}{*}{$\log{gf}$} \\
\cline{3-6}
 & centre $[\unit{\AA}]$ & Start & End & Start & End &  \\
\hline 
Mg I & 4702.9909 & 4702.00 & 4704.00 & 4680 & 4720 & -0.44 \\
Ti II & 4563.7574 & 4562.75 & 4564.75 & 4540 & 4580 & -0.69  \\
Ti II & 4571.9713 & 4570.97 & 4572.97 & 4550 & 4590 & -0.31  \\
Fe II & 4583.8292 & 4582.70 & 4584.70 & 4560 & 4600 & -1.86  \\
Ba II & 4554.0290 & 4553.00 & 4555.00 & 4535 & 4575 & 0.17 \\
\end{tabular}
\end{table*}

\subsection{Stellar parameter estimates}\label{sec_SP}
We estimated abundances using spectral fitting. This required some estimates of the stellar parameters effective temperature $\teff$, surface gravity $\logg$, and metallicity $\mh$ -- and in most cases also microturbulence $\vmic$. As described in Sect.~\ref{sec_algorithm}, we used a single broadening parameter to capture the combined effects of macroturbulence, stellar rotation, and instrumental broadening -- the last of which is expected to dominate. 

$\teff$ and $\logg$ were determined by projecting the stars onto Victoria isochrones~\citep{victoria}. In Paper VI, Stetson's $V$~magnitude and $V-I$~colour of the 150 stars studied in that article were used to fit a $13.4\,\unit{Gyr}$, $\mh = -2.3$ isochrone for M30. We show our estimated $\teff$, $\logg$, and $\mh$ (described below) in Fig.~\ref{ill_kiel}, together with the stellar isochrone used in model T$6.09$. There is a slight inconsistency between the isochrone used for the parameter estimates and that for the stellar evolution model. To the extent that this matters, it may introduce a slight tendency to underestimate the abundance difference between hot and cold stars. Paper~VI also compared Victoria isochrones to BaSTI~\citep{basti} and Dartmouth isochrones~\citep{dartmouth} and found that only the Victoria isochrones were capable of reproducing the observed morphology of the cluster~\citep[Sect. 3.1]{same_6}. In Appendix~\ref{sec_teff} we investigate the effect of a possible offset in $\teff$ and the resulting shifts in $\logg$ and $\vmic$.

The overall metallicity $\mh$ for each star was determined using spectral synthesis methods similar to those used in this paper. We note that $\mh$ is not crucial to our analysis: Our spectral synthesis code (described in Sect.~\ref{sec_sme}) requires some assumed overall metallicity, but the result of fitting an abundance to a specific line should ideally not depend on this assumption. The only way it should be able to affect our abundance estimates is if the lines we used to measure abundances are blended with lines of other elements, which in turn are included in our line list. In that case, the accuracy of the $\mh$ estimate affects how well we can compensate for the blend.

\begin{figure}
\centering
\includegraphics[width=8.8cm]{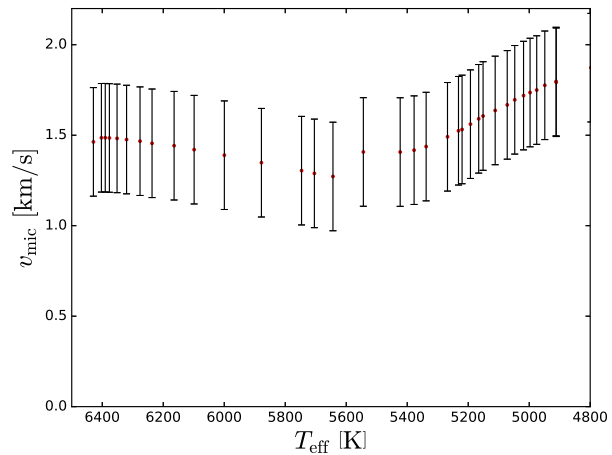}
\caption{Estimated microturbulence parameter $\vmic$ for the spectra used in this analysis. The error bars show the range over which each microturbulence was varied in the analysis, as described in Sect.~\ref{sec_systerror_source}.}
\label{ill_vmic}
\end{figure}

As described in Sect.~\ref{sec_nlte}, the microturbulence parameter $\vmic$ is necessary when using 1D models of the stellar atmosphere. Normally, $\vmic$ would be estimated using spectral fitting, but our spectra do not contain enough lines for this. The upcoming Paper~VIII in this article series, investigating a sample similar to this, finds that it was not possible to robustly estimate $\vmic$. A preliminary discussion of this is found in~\citet[Sect. 3.4.1]{same_8}. This forced us to use some empirically determined relation for estimating $\vmic$ from the known stellar parameters. Paper~VI in this article series used an empirical relation that is commonly used within Gaia-ESO~\citep[Sect. 3.3]{same_6}. The same formula is explicitly described in~\citet[Appendix E]{me_1}. However, it was calibrated on population~I stars, and is likely not appropriate for our current sample. Instead, for stars with $3.5 > \logg$ we used a relation estimated in~\citet{sitnova} based on dwarf stars with $-2.6 \leq \feh \leq 0.2$:
\begin{equation}
\frac{\vmicdwarf}{\left[ \unit{km/s} \right]} = 
-0.21 + 0.06 \cdot \feh + 5.6 \frac{\teff}{10^4 \unit{K}} - 0.43 \cdot \logg.
\label{eq_vmic_dwarf}
\end{equation}
For stars with $\logg < 3.5$ we used a relation estimated in~\citet{mashonkina} for giant stars with $-4.0 < \feh < -1.7$:
\begin{equation}
\frac{\vmicgiant}{\left[ \unit{km/s} \right]} = 
0.14 - 0.08 \cdot \feh + 4.9 \frac{\teff}{10^4 \unit{K}} - 0.47 \cdot \logg.
\label{eq_vmic_giant}
\end{equation}
As discussed in Sect.~\ref{sec_systerror_source}, this is expected to be one of the two dominant sources of systematic error, which we took into account by varying the $\vmic$ within $\pm 0.3\,\unit{km/s}$ around the values predicted by the formula. Figure~\ref{ill_vmic} shows the $\vmic$ for each spectrum used in the fitting, together with the range of variation. There is a discontinuity at the dwarf-giant boundary, but it is small compared to the variation in $\vmic$.

The spectra of individual TOP stars generally have too low $\snr$ to be reliably analysed, especially since we have found $\snr$-dependent systematic errors, which we discuss in Sect.~\ref{sec_systerror_source}. This means that we would prefer to have spectra with approximately equal $\snr$. As described in Appendix~\ref{app_combined}, we coadd spectra of stars with similar stellar parameters to produce group spectra with $\snr \approx 100$.

\subsection{External software}
The analysis pipeline uses several external pieces of software. The pipeline is built around the spectral synthesis code SME. SME in turn requires a line list and a grid of model atmospheres. We describe SME briefly in section~\ref{sec_sme}.

For the line list, we use the line list that has been compiled for use in the Gaia-ESO Survey, based on the best literature values available around 2012--2014, when the survey began~\citep{ges, linelist}. We use model atmospheres calculated with the Model Atmospheres with a Radiative and Convective Scheme (MARCS) code, under spherical symmetry and with $\vmic=2\,\unit{km/s}$ for stars with $\logg \leq 3.5\,\unit{dex}$, and with plane-parallel geometry and $\vmic=1\,\unit{km/s}$ for stars with $\logg > 3.5\,\unit{dex}$~\citep{marx}.

In addition, the pipeline makes use of the IDL Astronomy User's Library (AstroLib) and the Coyote IDL library~\citep{idlastro, coyote}. Outside of the pipeline proper, several Python modules were used: Astropy was used to handle spectral files~\citep{astropy_1, astropy_2}. SciPy was used to handle most calculations~\citep{scipy}. emcee was used to run Goodman and Weare's Affine Invariant Markov Chain Monte Carlo Ensemble Sampler for model fits~\citep{affine_invariant, emcee}. Matplotlib was used to make all plots in this article~\citep{matplot}.

\subsubsection{SME}\label{sec_sme}
SME is a code for synthesising and fitting stellar spectra. The earliest published version is described in~\citet{smeoriginal} and the most important changes since then are documented in~\citet{smevolution}. Our pipeline uses SME version~542.

SME calculates synthetic spectra on the fly, based on a line list and a grid of model atmospheres. To determine stellar parameters and/or abundances, it fits synthetic spectra to an observed spectrum~\citep{smevolution}. The fitting method resembles a $\chi^2$-minimisation, but tweaked to be more robust to the line list being incomplete~(Piskunov, priv. comm). When estimating abundances, SME also provides estimates of the uncertainties on the estimates, as described in Sect.~\ref{sec_randerror}.

We determined line-by-line abundances. This means that for each line, we defined a segment mask and a line mask. The segment mask simply specifies which pixels in the observed spectrum should be synthesised at all. The line mask specifies which pixels should be used in the fitting itself. We show the segment and line masks for each line in Table~\ref{table_lines_sme}.

Before the abundance determination, the segment is normalised using a linear rescaling that tries to fit the observed spectrum to an initial model spectrum, which is calculated using the first guess of the abundances. Current versions of SME allow doing this by a weighted least-squares minimisation over all pixels in a segment~(Piskunov, priv. comm). Older versions instead require the pixels used in the fitting to be explicitly assigned to a continuum mask, and only use an unweighted least-squares minimisation~\citep[Sect. 3.4]{smeoriginal}. We use the newer method in the main body of the article, but since -- as discussed in Sect.~\ref{sec_systerror_source} -- the continuum normalisation is expected to be one of our two main sources of systematic error, we compare to the results of using the older method in Appendix~\ref{app_cont}.

In SME, the abundances in stars are by default assumed to follow the Solar abundance pattern as estimated by~\citet{Grevesse_Asplund_Sauval}, but rescaled by $\mh$. In addition, in our pipeline, the abundances of the $\alpha$~elements O, Ne, Mg, Si, S, Ar, Ca, and Ti are increased by $0.4$, which is an empirical estimate of $\afe$ at this metallicity.

\subsubsection{Approximations in radiation transport}\label{sec_nlte}
By default, SME assumes that the stellar atmosphere is 1D and that radiation transport occurs in local thermodynamic equilibrium (LTE), meaning that atomic level populations can be estimated using Saha-Boltzmann statistics. In some cases we can take into account non-LTE (NLTE) effects, and in some cases we can at least partly take 3D effects into account. As a rule, we cannot do both at once. In our analysis we used 1D~LTE for Ti and Fe, and 1D~NLTE for Mg and Ba. However, in Appendix~\ref{app_3D_comp} we consider indicative 3D~LTE calculations for Ti and Fe.

The 1D approximation is expected to be poor at describing convection near the optical surface since the interaction between convective motion and radiative energy transport creates both temperature inhomogeneities across the surface -- which cannot be described in 1D -- and a steeper temperature gradient than predicted in 1D. There are also local velocity fields that give rise to a desaturation effect, which we model in 1D as microturbulence with the parameter $\vmic$. All these cause the Curves of Growth (COG) for saturated lines -- such as the Ti~II and Fe~II lines in red giants -- to be much more accurate in 3D than 1D. On the other hand, the overionisation of Mg~I at these metallicities means that the LTE approximation is likely to get worse in 3D since the steeper temperature gradient increases the NLTE effects.

The LTE assumption is expected to give good results for the Ti~II and Fe~II lines, but not for the Mg~I and Ba~II lines. The stars in M30 are very metal-poor, which means that the lack of metal-line opacities causes the local gas and radiation field to decouple. This results in overionisation, meaning that the neutral ground state is depleted relative to the predictions of LTE. The singly ionised majority species -- Ti~II and Fe~II -- are relatively immune to this since they have higher ionisation potentials.

For the NLTE computation, we used a grid of departure coefficients that estimate how much the level populations at different depths in the atmosphere are expected to differ between NLTE and LTE. The capacity for using such grids is built into SME~\citep[Sect. 3]{smevolution}. We used two pre-calculated grids, one for Ba and one for Mg. %We also avoid the line regions that still cannot be synthesised well by cutting out pixels whose normalised flux drops below~$0.2$.
For Mg, we used a grid described in \citet{anishs_new_article}, using a model atom described in \citet{modelatom}.

We performed 3D~LTE spectrum synthesis calculations using SCATE~\citep{scate} on 3D hydrodynamic model atmospheres from the STAGGER grid~\citep{stagger}. We sampled the temporal evolution by using 20 snapshots, selected at regular intervals, from each hydrodynamic simulation. For comparison, 1D~LTE radiative transfer calculations were performed using the same codes on model atmospheres computed with the 1D hydrostatic model atmosphere code ATMO~\citep{stagger} which implements the same microphysics as STAGGER. The radiative transfer calculations were performed over a range of abundances in steps of $0.1\,\unit{dex}$. 3D~LTE-1D~LTE abundance corrections were derived by interpolating the 3D~LTE equivalent widths onto the 1D~LTE curve of growth. 

\section{Model specification}\label{sec_model_spec}
% Explain what we want to do
Our goal is to constrain $T_0$, assuming the stellar evolution models described in Sect.~\ref{sec_trend}. We also compared our results to a null model that assumes that abundances do not vary with stellar parameters at all, to test whether we could rule out the possibility that diffusion processes have no appreciable effect on abundances.% Note that, strictly speaking, this null model is equivalent to a stellar evolution model with $T_0$ set very high.

It would seem to be simple to compare these models: We could simply fit the derived abundances to the predictions of all models, and then compare the quality of the best fits using the metric of our choice. Unfortunately, things are more complicated than they seem: Both the derived abundances and model predictions have systematic uncertainties of comparable size to the random uncertainties. This means that we cannot simply calculate some goodness-of-fit metric for each value of $T_0$ and choose the best one. We describe the origins of the systematic uncertainties in Sect.~\ref{sec_systerror_source} and our method for constraining them in Sect.~\ref{sec_systerr_constrain}. In Sect.~\ref{sec_randerror} we state our method for estimating statistical uncertainties. In Sect.~\ref{sec_formalities} we state the likelihood function that follows from the assumptions made. In Sect.~\ref{sec_null} we describe the null hypothesis that we used to verify that our results look noticeably different from what would be expected if diffusion processes are inefficient.

\subsection{Sources of systematic error}\label{sec_systerror_source}
% Explain why it is hard to compare them
In stellar spectroscopy, derived parameters such as elemental abundances often have systematic uncertainties of comparable size to statistical uncertainties. In this specific case, we know that the abundances derived by our pipeline have systematic errors that depend strongly on $\teff$ and $\logg$. In addition, the derived abundances depend systematically on the $\snr$, which for observational reasons depend on $\teff$ and $\logg$ as well. Before we could make any well-motivated statement about how well the derived abundances fit any hypothesis, we needed to compensate for or at least constrain these systematic effects. In addition, an even bigger source of systematic uncertainty is that some starting abundance for each element must be assumed in the stellar evolution model, before the evolution over time can be calculated. However, unless the assumed value is very far off from the true value, this simply creates a $\teff$-independent constant offset in the (logarithmic) abundances.

% Explain how we think the systematic errors work
A general overview of sources of systematic uncertainty in stellar spectroscopy can be found in~\citet{accuracy}. In this study we assumed that the dominant sources of systematic uncertainty are the continuum placement and the choice of $\vmic$.
% Explain continuum level
As described in Sec.~\ref{sec_algorithm}, it is necessary to estimate the continuum level around a spectral line before the line can be used to estimate an elemental abundance. This estimate is never perfect, and relatively small errors can have a large impact on derived abundances. If this simply led to a random scatter in the derived abundances, it would be easy to deal with, but unfortunately it is likely to depend on the stellar parameters. A detailed study supporting this assumption can be found in~\citet{black_box}.
% Explain vmic
As described in Sect.~\ref{sec_SP} the microturbulence parameter $\vmic$ must be estimated for those abundance estimations that use a one-dimensional model of the stellar atmosphere, and we have no way of directly measuring it. There are many methods for estimating $\vmic$ based on $\teff$, and any one of them will systematically differ from the true $\teff$ dependence, imposing a systematic error in the derived abundances.

% Explain why we can sort of deal with other sources of error as well
In principle, there are many other sources of systematic error, due to both observational biases and imperfections in the modelling. However, we to some extent implicitly take them into account, as long as they are much smaller than the two sources of systematic errors that we explicitly take into account. Our method for compensating for the systematic errors only requires some upper bound on their size, while being otherwise agnostic about what the actual source of the errors is. Hence, as long as the additional sources of error are small enough to fit within the bounds set by those two sources of error, it can compensate for them.
% Example about weak lines
As an example of an additional source of systematic error, \citet[Sect.~3.6 \& Fig. 6]{same_5} -- Paper~V in this article series -- found that this type of study is also affected by systematic errors in abundances derived from weak lines. The underlying reason is that the error estimation described in Sect.~\ref{sec_randerror} implicitly assumes that the expected flux depends linearly on the parameter being estimated. However, while the expected flux is approximately linear in the relative number of absorbers $\nx / \nh$, we define abundances on a logarithmic scale. For weak lines the pixel noise is proportionally large compared to the change in expected flux given by a change in the estimated abundance, which creates a non-negligible asymmetry in the error on the abundance estimate.

% Andreas' suggested comment is:
% The two main sources of systematic effects considered here (position of the continuum and microturbulence value) can be seen as placeholders for observational and modelling biases, respectively. Echelle spectra can usually not be continuum normalized to arbitrary levels of accuracy because they are not entirely free of instrumental and observational effects (see e.g. Korn 2001). Likewise, a $T-\tau$ relation with too shallow a $T$ gradient (as is the case for 1D models of metal-poor stars) will result in systematically larger microturbulence values (Gray 2005). So while we focus on these specific effects for implementation reasons, they can be taken to represent general observational, instrumental and modelling shortcomings. 

% Explain how we deal with systematic errors
\subsection{Method for constraining systematic uncertainties} \label{sec_systerr_constrain}
To constrain the systematic uncertainties, we need to make six assumptions. The first assumption is that for a given element, the arbitrariness in the assumed abundances of the primordial gas in the stellar evolution models will affect model abundances in all stars with a scaling factor that is constant over the cluster. The second assumption is that the systematic error not due to the initial abundance choice is due to offsets in the continuum level and arbitrariness in the assumed $\vmic$. The third assumption is that by looking at a spectrum, a human spectroscopist can give a reliable estimate of what range the correct continuum level could plausibly lie within. The fourth assumption is that for a given spectrum, the component of the error that is uncorrelated from spectrum to spectrum will be approximately the same no matter how the continuum is placed for that particular spectrum. The fifth assumption is that for a sample of stars such as ours, $\vmic$ is somewhere within $\pm 0.3\,\unit{km/s}$ of those predicted by Eqs.~\eqref{eq_vmic_dwarf} and~\eqref{eq_vmic_giant}, and within this interval the derived abundance depends monotonically on $\vmic$. The sixth assumption is that the systematic error does not vary very quickly over the range of stellar parameters present in our sample, so that it can be well approximated with a low-order polynomial.
    
Assumption one means that we can add an arbitrary constant to the log-abundances for each element. In practise, we shifted the average abundance of the stars below $5500\,\unit{K}$ to have the same mean abundance as predicted by the model. Below that threshold the models make practically identical predictions. Assumptions two through five allowed us to use the following method for constraining the systematic uncertainty for each abundance estimate: We visually inspected the fitted spectra for each abundance, and noted within which range the continuum could reasonably vary. Doing this, we found that the continuum must be somewhere within $0.5\%$ of SME's best estimate. We re-derived abundances with the continua shifted to the top and bottom of that range. We also re-derived abundances with $\vmic$ varied by $0.3\,\unit{km/s}$, giving us nine derived abundances in total. The exception is in Appendix~\ref{app_3D_comp}, where the 3D-modelling makes the $\vmic$ parameter unnecessary. We denote the highest and lowest estimates of the abundance by $\abmax$ and $\abmin$, respectively. Assumption six then allows us to model the systematic error $e_i^\mathrm{syst}$ as a polynomial subject to the constraint that for each spectrum $\left| \esyst \right| \leq \abmax - \abmin$. Since we do not know how quickly the systematic error may vary, we modelled it with a range of polynomials of increasing order. In Sect.~\ref{sec_aic} we describe how we dealt with the problem that a higher-order polynomial necessarily gives a better fit.

% Explain how our hypotheses work
When fitting the stellar evolution models described in Sect.~\ref{sec_trend}, we assumed that the derived stellar abundances follow normal distributions centred around the values predicted by the models, plus the systematic offsets. Sect.~\ref{sec_randerror} describes how the width of the distributions was estimated.

% Explain how we deal with systematic errors dependent on S/N
While the dependence of derived abundances on $\snr$ can in principle be handled in the same way as the dependences directly on the stellar parameters, we used a simple method to mostly cancel it out: Instead of deriving abundances for the spectra of individual stars, we coadded spectra with similar stellar parameters together, in such a way that the grouped spectra have approximately the same $\snr$. Given this, this source of systematic error should to first approximation affect all derived abundances equally, which makes it easier to take into account. We show this coaddition in Appendix~\ref{app_combined}.

% This should come after the discussion of the systematic errors, to emphasise that it's actually of lesser importance.
\subsection{Method for estimating statistical uncertainties}\label{sec_randerror}
We had two sources of discrepancies between our derived and predicted abundances that have to be treated as statistical errors. First of all, pixel noise in our spectra created some amount of scatter in our derived abundances. Secondly, there is some intrinsic random scatter in the true abundances in the stars, which is physically real but uninteresting for our attempts at fitting abundance trends.

Current versions of SME estimate statistical errors using a heuristic algorithm described in~\citet{smevolution}. That algorithm was derived on the assumption that multiple lines are being fitted, and attempts to take into account systematic shifts in the abundance scale due to uncertain line parameters. It is probably not appropriate for estimating the errors in single-line abundances. To get meaningful estimates of the statistical errors, we restored the older error estimation method in our copy of SME. This takes the errors directly from the covariance matrix generated in the fitting, as described in~\citet{smeoriginal}. This essentially assumed that our statistical errors come entirely from pixel noise. We also imposed a lower bound of $0.05\,\unit{dex}$ on the errors, since we do not believe any error estimate lower than that is realistic.

We assumed that the intrinsic scatter in abundances is normally distributed with a standard deviation $\sint$ that is specific to each element. We also made the simplifying assumption that within the spectra used to create our group spectra, as described in Sect.~\ref{app_combined}, the variation in abundance is small enough to have an approximately linear effect on the abundances -- meaning that in our spectral fitting the group spectra could be expected to behave as though they had abundances with variations $\sint / \sqrt{\nstar}$, where $\nstar$ is the number of stars included in each group spectrum.

Magnesium should have the greatest scatter, since it is affected by anticorrelations. Paper~II in this series found intrinsic scatter with a magnitude somewhere below~$0.1\,\unit{dex}$ in the cluster NGC~6397~\citep[Sect. 4.2, Fig. 7]{same_2}. Paper~V got similar results for the cluster NGC~6752~\citep[Fig. 3]{same_5}. Based on this, we adopted $0.1$ as a conservative estimate of $\sintmg$. Fe and Ti are not affected by anticorrelations, but still have some observed scatter~\citep[Sect 3.3]{chromosomes_2017}. For Fe, an upper limit of $0.05\,\unit{dex}$ has been established~\citep{intrinsic_iron}, so we adopted this as a conservative estimate of $\sintfe$. Since Ti should behave similarly to Fe, we adopted this as our estimate of $\sintti$ as well.

\subsection{Likelihood function}\label{sec_formalities}
We modelled the error by assuming that for each star $i$ the derived abundance $\ab$ is given by
\begin{align}
\ab &= \abtrend + \aoffset + \abint + \esyst + \erand, \label{eq_ab}
\end{align}
where $\abtrend$ is the abundance predicted by the model, whether the null model or diffusion, $\aoffset$ is some number that does not vary from star to star, $\abint$ is a normal distribution with mean zero and standard deviation $\sint$, $\esyst$ is a polynomial in $\teff$ and $\erand$ is a normal distribution with mean zero and standard deviation $\si$ given by SME's uncertainty estimates as described in Sect.~\ref{sec_randerror}.

The likelihood for a particular $\ab$ and element X is then given by
\begin{align}
L \left( \ab \right) &=
\left\{
\begin{aligned} \frac{\exp{\left( - \frac{ \left( \abtrend + \esyst - \ab \right) }{2 \left( \si^2 + \sint^2 \right)} \right)}}{\sqrt{2 \pi \left( \si^2 + \sint^2 \right)}}, &\hspace{3mm} \ab + \esyst \in \left[ \abmin, \abmax \right] \\
0, &\hspace{3mm} \ab + \esyst \notin \left[ \abmin, \abmax \right] .
\end{aligned}
\right.
\end{align}
Since the $\ab$ are uncorrelated (by construction, all correlations are built into trends and systematics) the likelihood for the full sequence of estimates $\left\{ \ab \right\}$ is given by
\begin{align}
L \left( \left\{ \ab \right\} \right) &= \prod_{i=0}^n L \left( \ab \right). \label{eq_L_ab_all}
\end{align}
For each line we calculated best fits to several models of the form \eqref{eq_ab}. We used the T$5.8$, T$5.9$, T$5.95$, T$6.0$, T$6.09$, T$6.2$, and null models to describe the true trend $\abtrend$ and a zeroth to fifth degree polynomial in $\teff$ to describe the systematic error $\esyst$, making $7 \cdot 6 = 42$ models in total. We did not go beyond that since we expect an error model with six free parameters to be able to fit practically any data.

\subsection{Definition of null hypothesis} \label{sec_null}
% Explain how we define our null hypothesis
We took as our null hypothesis that the abundances follow a normal distribution independent of the stellar parameters. As the mean of the distribution, we took the solar abundance scaled by the metallicity and expected $\left[ \alpha/\mathrm{M} \right]$ of the M30 stars. As the standard deviation of the distribution we used the same $\sint$ as for the stellar evolution models.

\section{Model comparison}
Given the likelihood function \eqref{eq_L_ab_all}, each combination of diffusion trend and degree of the systematic error polynomial has some maximum likelihood $\hat{L}$. We want to turn those likelihoods into some quantitative statements about how likely the different values of $T_0$ are, and also how well our data could be distinguished from what we would see under the null hypothesis that diffusion is inefficient. We compare all models using the corrected Akaike information criterion~($\aicc$), described in Sect.~\ref{sec_aic}. We also compare the best-fit stellar evolution model to the null hypothesis by using the likelihood ratio, described in Sect.~\ref{sec_likelihoodratio}. We make a caveat about the `garden of forking paths' issue with the statistical hypothesis testing in Sect.~\ref{eq_forks}.

\subsection{Akaike information criterion}\label{sec_aic}
Formally, the $\aicc$ for a model is defined as
\begin{equation}
    \aicc = 2 k + \frac{ 2 k \left( k + 1 \right) }{n - k- 1} - 2 \ln{\hat{L}}, \label{eq_AICc}
\end{equation}
where $n$ is the number of data points, $k$ is the number of free parameters in the model and $\hat{L}$ is the maximum likelihood of the data -- that is, the likelihood given the best-fitting model parameters, which in our case are $T_0$ and the coefficients of the polynomial describing the systematic error. (Some authors choose to drop the factors of $2$ from the definition). In a set of models, $\daicc$ for a model is the difference between the $\aicc$ for that model and that of the model with the lowest $\aicc$. When presenting results, we will only show $\daicc$.

A detailed discussion of the definition and interpretation of the $\aicc$can be found in~\citet{Burnham_Anderson} or~\citet{AIC_BIC}. 
% I think we definitely should say or, since there's not point in reading both. They're a book and an article that contains the most important points in the book.
However, we attempt to give an intuitive explanation here: None of the statistical models that we were testing is exactly true, but we wanted to identify which one is closest to the truth. One measure of how `close to the truth' a statistical distribution is, is the Kullback-Leibler divergence (KL divergence) between that distribution and the true distribution. Since we do not know the true distribution, we cannot calculate this directly. However, by looking at the difference $\daicc$ between two models, we can estimate the probability that one model has a lower KL divergence from the truth than the other.

Translating $\daicc$ into these probabilities is not completely straightforward. As a heuristic, if two models differ by $\daicc < 2$, they have comparable probabilities of being the KL-minimising model, and if they differ by $\daicc > 10$, the one with lower $\aicc$ is virtually certain to be KL-minimising. This can be made more rigorous by introducing Akaike weights, defined as
\begin{align}
    w_i &= \frac{\exp{\left( - \frac{1}{2} \daicc^i \right)}}{ \sum_r \exp{\left( - \frac{1}{2} \daicc^r \right)} }, \label{eq_akaike_weight}
\end{align}
where the summation is performed over all $\daicc$. These weights can be seen as Bayesian posteriors for model $i$ being KL-minimising.

One advantage of this method is that it neither requires nor prevents us from adopting a Bayesian framework. Contrary to common misconception, the AIC and its corrections can be derived on either frequentist or Bayesian grounds. In addition, since it looks at which statistical model is likely to be closest to the true distribution this method does not assume that either model is true. The disadvantage is that we are measuring `closeness' in terms of information loss. While this would be the correct choice if we were only interested in estimating out-of-sample uncertainty -- that is, if we only wanted to decide which model to use when predicting abundances in the stars of M30 that we have not yet looked at -- it is not clear that it is the best measure of which of the underlying physical hypotheses is likely to be true.

Using the likelihoods \eqref{eq_L_ab_all} for each of the 30 model fits we calculated $\aicc$ for each model. Since the $\aicc$ is additive -- as can be seen from \eqref{eq_AICc} -- we looked at the best fit over all lines, but we also looked at the $\aicc$ for the individual lines separately. We could in principle look at combinations of the error models, such as a linear error in Ti but quadratic in Fe, but we do not do so and instead assume that if the error in one line must be described with a polynomial of a certain degree, it must be for all lines.

\subsection{Likelihood ratio}\label{sec_likelihoodratio}
Aside from the use of the Akaike information criterion to identify the support for different models, we also looked at the likelihood ratios between the diffusion and null models. This to allow us to demonstrate that diffusion processes do have effects that can be distinguished from systematic errors, irrespective of how well $T_0$ can be constrained. When comparing the null and stellar evolution models, we took the likelihoods for the stellar evolution models identified as the best, in the sense of having the lowest $\aicc$. Since the likelihood is monotonously increasing with the number of parameters describing the systematic error, there is no unique highest-likelihood model.

The mathematical background for this section can be found in~\citet{FPR}. The likelihood ratio is formally defined as
\begin{align}
L_{10} &= \frac{\hat{L}_\mathrm{Add\-Mix}}{\hat{L}_\mathrm{null}}.
\end{align}
Using this, we could calculate the False Positive Rate (FPR). This is a Bayesian quantity, stating the estimated risk of falsely accepting the alternative hypothesis, given some prior probability that it is true. This is a useful metric since it essentially corresponds to what readers often incorrectly take the $p$-value to mean, implying that it expresses what most readers looking at $p$-values actually want to know. (A detailed discussion of common misconceptions of $p$-values can be found in~\citet{dirty_dozen}).
% I'm not going to put this in the text, but note that article V in this article series makes exactly this mistake.
Formally, the FPR is defined as
\begin{align}
\FPR &\equiv \frac{1}{1 + L_{10} \frac{\padd}{1 - \padd }}, \label{eq_FPR}
\end{align}
where $\padd$ is the prior probability that the diffusion hypothesis is true.

% Explain FPR
If we accept a prior of $0.5$ for both the null hypothesis and the diffusion hypothesis, this simplifies to
\begin{align}
\FPR &= \frac{1}{1 + L_{10}}. \label{eq_FPR_even}
\end{align}
% Explain reverse bayesianism
Alternatively, we can accept a $\FPRmax = 0.05$, and calculate the smallest prior $\padd$ that one could have that would still allow us to accept the hypothesis:
\begin{align}
\padd &= \frac{1 - \FPR_\mathrm{max}}{1 + \left( L_{10} - 1 \right) \FPR_\mathrm{max}}. \label{eq_revbayes}
\end{align}

One advantage of using the likelihood ratio is that it gives us quantitative results that have fairly straightforward interpretations. One disadvantage is that we have to select either $\padd$ or $\FPR$ more or less arbitrarily. A second disadvantage is that it implicitly assumes that one or the other of our models is true: In reality, while one or the other is true of the physical hypotheses that diffusion does or does not appreciably affect stellar spectra, the model hypotheses that the stellar abundances are sampled from statistical distributions of the form~\eqref{eq_ab} are certainly all false. Whatever distribution reality is sampling from, it is guaranteed to be more complex than our models.

\subsection{A caveat about forking paths}\label{eq_forks}
We should point out that neither the values of $\daicc$ or $\FPR$ can be taken entirely at face value. Strictly speaking, statistical measures of this kind can only be interpreted in a straightforward way when the procedure for statistical analysis was decided on before the data were seen. When the analysis was adapted in response to the data, the study has to be seen as exploratory rather than confirmatory, and statistical measures have to be treated as heuristics. This issue is discussed in~\citet{forks} regarding $p$-values specifically, but the argument is general enough to apply to any statistical measure.

In our case, several important discoveries were made partway through the statistical analysis. These include, but were not limited to, finding that there are non-negligible parameter-dependent systematic uncertainties, that spectra of individual stars tend to have too low $\snr$ to be useful, that systematics also depend on $\snr$ and $\vmic$, and that there are two very different but arguably reasonable methods for calculating SME uncertainties. Hence, we have to consider this an exploratory study.

\section{Results and discussion}
We estimate abundances for our sample of group spectra of stars from the TOP to the tip of the red giant branch (RGB) of M30, as described in Sect.~\ref{sec_abundance_derivation}. We use those abundances to fit statistical models that contain both the predictions of stellar evolution models and a model of parameter-dependent systematic errors, as described in Sect.~\ref{sec_model_spec}. We then compare the models using the Akaike information criterion as a way of fixing the parameter $T_0$ that describes the efficiency of Add\-Mix processes in the stellar evolution models, as described in Sect.~\ref{sec_aic}. We also use $\aicc$ and likelihood ratios to compare the models to the null hypothesis that there are no overall trends -- the likelihood ratios being described in Sect.~\ref{sec_likelihoodratio}.

We discuss the results of using all lines together in Sect.~\ref{sec_res_all} We discuss the results for Mg, Ti and Fe individually in Sects.~\ref{sec_res_Mg}, \ref{sec_res_Ti}, and \ref{sec_res_Fe}. We make some concluding remarks in Sect.~\ref{sec_res_conc}.

\subsection{All lines}\label{sec_res_all}
We show all derived abundances in Table~\ref{table_big_spectable}, together with other properties of each spectrum. In Table~\ref{table_full_spectype_Combined_LOGG_mean5500_NLTE_new_vmic_constraining_systematics_error_floor_0.100} we show the combined $\daicc$ for all lines, assuming that the true trend is either that predicted by one of our stellar evolution models, or simply flat, and that the systematic error is somewhere between constant and a 5th degree polynomial.

Just by going with the heuristic that models whose $\daicc$-values differ by less than $2$ are more or less indistinguishable, while models that have $\daicc$ larger than $10$ can probably be ruled out, we can draw the conclusion that the models closest to reality are almost certainly T$6.09$ and T$6.2$, and the null model can be ruled out. Our systematic errors are probably best modelled as either constant or a linear offset in $\teff$, although the second- and third-order polynomials are still plausible. No matter how we model our systematic errors, T$5.8$, T$5.9$, T$5.95$, and T$6.0$ are definitely ruled out. 

We show the Akaike weights -- as defined in \eqref{eq_akaike_weight} -- for the  models in Table~\ref{table_bayes_spectype_Combined_LOGG_mean5500_NLTE_new_vmic_constraining_systematics_error_floor_0.100}. Within a Bayesian framework, the weights can be interpreted as posterior probabilities of each model being KL-minimising, but even within a Bayesian framework the caveat in Sect.~\ref{eq_forks} still applies. The weights say essentially the same as the rule-of-thumb interpretations above. The probability is negligible that any model other than T$6.09$ or T$6.2$ is KL-minimising.

We show the log-likelihoods of the linear-offset models in Table~\ref{table_full_logL_spectype_Combined_LOGG_mean5500_NLTE_new_vmic_constraining_systematics_error_floor_0.100}. With these, we can compare the best-fitting null model (with linear systematic error) to the best-fitting stellar evolution model ($T6.2$ with constant systematic error) outside of the $\aicc$ framework. According to \eqref{eq_FPR_even} rejecting the null hypothesis based on this would give us a $\FPR$ of $2.56 \cdot 10^{-13}$. Conversely, \eqref{eq_revbayes} states that we should be willing to proceed on the assumption that diffusion processes are real as long as our prior was above $4.88 \cdot 10^{-12}$.

In summary, irrespective of what framework we use, we can conclude that $\log{\left( T_0 / \left[ \unit{K} \right] \right)}$ is probably somewhere around $6.09$-$6.2$. We also find that in light of the data, we can rule out the null hypothesis that the abundance trends are purely an effect of systematics. The AddMix efficiency T$6.09$-$6.2$ derived here is somewhat higher than that found by Papers~I-III in this article series when they investigated the cluster NGC~6397, which is a factor of~two more metal-rich than M30~\citep{same_1, same_2, same_3}.

Our findings mostly support the conclusions of Paper~VI. That article attempted to investigate the Spite plateau by estimating abundances for lithium and several elements in stars in M30. They used T$6.0$ in their analysis, based on Paper~III, but tentatively noted that T$5.8$ actually gave a better fit to their derived calcium abundances. Based purely on our findings, T$6.0$ was the best choice of the two, but either T$6.09$ or T$6.2$ could have been better.

\begin{table}
\caption{$\daicc$ values for all four lines of Fe, Ti, and Mg together. Each column assumes a particular stellar evolution model, as described in Sect.~\ref{sec_trend}, and each row assumes a particular model of the error, as described in Sect.~\ref{sec_systerr_constrain}. Values above $10$ are written in grey, since they represent models that can essentially be ruled out.}\label{table_full_spectype_Combined_LOGG_mean5500_NLTE_new_vmic_constraining_systematics_error_floor_0.100}
\centering
\begin{tabular}{ l | c | c | c | c | c | c | c }
$T_0$: & 5.8 & 5.9 & 5.95 & 6.0 & 6.09 & 6.2 & null \\
\hline 
const. & $\textcolor{gray}{180}$ & $\textcolor{gray}{64.8}$ & $\textcolor{gray}{35.6}$ & $\textcolor{gray}{16.9}$ & $2.59$ & $0.00$ & $\textcolor{gray}{58.0}$ \\
\hline 
lin. & $\textcolor{gray}{134}$ & $\textcolor{gray}{46.6}$ & $\textcolor{gray}{25.4}$ & $\textcolor{gray}{10.7}$ & $0.97$ & $2.41$ & $\textcolor{gray}{24.4}$ \\
\hline 
quad. & $\textcolor{gray}{136}$ & $\textcolor{gray}{52.0}$ & $\textcolor{gray}{31.7}$ & $\textcolor{gray}{17.3}$ & $6.60$ & $7.42$ & $\textcolor{gray}{22.3}$ \\
\hline 
cubic & $\textcolor{gray}{135}$ & $\textcolor{gray}{51.1}$ & $\textcolor{gray}{31.1}$ & $\textcolor{gray}{17.1}$ & $7.62$ & $8.39$ & $\textcolor{gray}{25.6}$ \\
\hline 
4th & $\textcolor{gray}{142}$ & $\textcolor{gray}{59.6}$ & $\textcolor{gray}{38.8}$ & $\textcolor{gray}{24.4}$ & $\textcolor{gray}{14.8}$ & $\textcolor{gray}{14.5}$ & $\textcolor{gray}{32.8}$ \\
\hline 
5th & $\textcolor{gray}{149}$ & $\textcolor{gray}{68.2}$ & $\textcolor{gray}{47.0}$ & $\textcolor{gray}{32.0}$ & $\textcolor{gray}{20.5}$ & $\textcolor{gray}{19.3}$ & $\textcolor{gray}{39.8}$ \\
\end{tabular}
\end{table}

\begin{table}
\caption{Akaike weight $w_i$, as defined by Eq.~\eqref{eq_akaike_weight}, for all lines combined. Weights below $0.01$ are written in grey, since they represent models that, under the Bayesian interpretation of Akaike weights, have a probability below $1\%$ of being KL-minimising.}\label{table_bayes_spectype_Combined_LOGG_mean5500_NLTE_new_vmic_constraining_systematics_error_floor_0.100}
\centering
\begin{tabular}{ l | c | c | c | c | c | c | c }
$T_0$: & 5.8 & 5.9 & 5.95 & 6.0 & 6.09 & 6.2 & null \\
\hline 
const. & $\textcolor{gray}{0.00}$ & $\textcolor{gray}{0.00}$ & $\textcolor{gray}{0.00}$ & $\textcolor{gray}{0.00}$ & $0.12$ & $0.44$ & $\textcolor{gray}{0.00}$ \\
\hline 
lin. & $\textcolor{gray}{0.00}$ & $\textcolor{gray}{0.00}$ & $\textcolor{gray}{0.00}$ & $\textcolor{gray}{0.00}$ & $0.27$ & $0.13$ & $\textcolor{gray}{0.00}$ \\
\hline 
quad. & $\textcolor{gray}{0.00}$ & $\textcolor{gray}{0.00}$ & $\textcolor{gray}{0.00}$ & $\textcolor{gray}{0.00}$ & $0.02$ & $0.01$ & $\textcolor{gray}{0.00}$ \\
\hline 
cubic & $\textcolor{gray}{0.00}$ & $\textcolor{gray}{0.00}$ & $\textcolor{gray}{0.00}$ & $\textcolor{gray}{0.00}$ & $\textcolor{gray}{0.01}$ & $\textcolor{gray}{0.01}$ & $\textcolor{gray}{0.00}$ \\
\hline 
4th & $\textcolor{gray}{0.00}$ & $\textcolor{gray}{0.00}$ & $\textcolor{gray}{0.00}$ & $\textcolor{gray}{0.00}$ & $\textcolor{gray}{0.00}$ & $\textcolor{gray}{0.00}$ & $\textcolor{gray}{0.00}$ \\
\hline 
5th & $\textcolor{gray}{0.00}$ & $\textcolor{gray}{0.00}$ & $\textcolor{gray}{0.00}$ & $\textcolor{gray}{0.00}$ & $\textcolor{gray}{0.00}$ & $\textcolor{gray}{0.00}$ & $\textcolor{gray}{0.00}$ \\
\end{tabular}
\end{table}

\begin{table}
\caption{Maximum (best-fit) log-likelihoods $\ln{\hat{L}}$ for all lines combined.}\label{table_full_logL_spectype_Combined_LOGG_mean5500_NLTE_new_vmic_constraining_systematics_error_floor_0.100}
\centering
\begin{tabular}{ l | c | c | c | c | c | c | c }
$T_0$: & 5.8 & 5.9 & 5.95 & 6.0 & 6.09 & 6.2 & null \\
\hline 
const. & $44.6$ & $102$ & $117$ & $126$ & $133$ & $134$ & $105$ \\
\hline 
lin. & $70.9$ & $115$ & $125$ & $133$ & $137$ & $137$ & $126$ \\
\hline 
quad. & $73.7$ & $116$ & $126$ & $133$ & $139$ & $138$ & $131$ \\
\hline 
cubic & $78.3$ & $120$ & $130$ & $137$ & $142$ & $142$ & $133$ \\
\hline 
4th & $79.4$ & $121$ & $131$ & $138$ & $143$ & $143$ & $134$ \\
\hline 
5th & $80.9$ & $121$ & $132$ & $139$ & $145$ & $146$ & $135$ \\
\end{tabular}
\end{table}

\subsection{Magnesium}\label{sec_res_Mg}
\begin{figure}
\centering
\includegraphics[width=8.8cm]{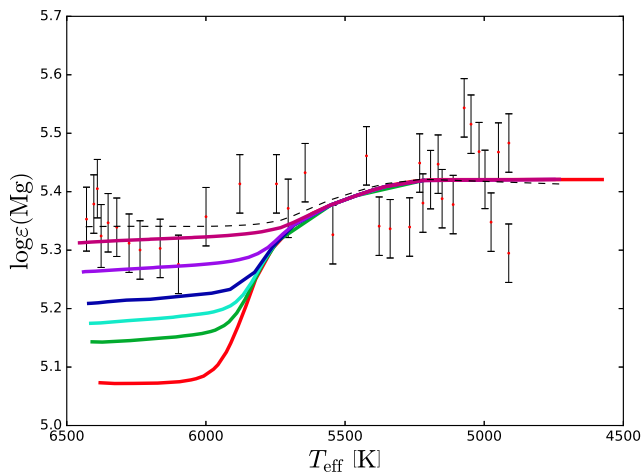}
\caption{Model predictions and estimated abundances for the Mg4702 line. The red curve shows the abundance predictions from T$5.8$, the green those from T$5.9$, the light blue those from T$5.95$, the dark blue those from T$6.0$, the violet those from T$6.09$ and the magenta those from T$6.2$. Red dots denote estimated abundances for each group spectrum. The black error bars show the random component of the error. The dashed black curve shows our estimate of the true trend with the best-estimate of the systematic error added.}
\label{ill_Mg4702_Full_model_fit}
\end{figure}

In Table~\ref{table_line_Mg4702_spectype_Combined_LOGG_NLTE_mean5500_new_vmic_constraining_systematics_error_floor_0.100} we show $\daicc$ for the fits to the Mg4702-line. The line clearly favours the null model. However, if it has to choose between one of the other models, it prefers T$6.2$, without showing any strong preference for any model of the systematic error. In Fig.~\ref{ill_Mg4702_Full_model_fit} we show the abundances as functions of $\teff$, together with the model predictions and the bounds on the systematic errors.

\begin{figure}
\centering
\includegraphics[width=6.6cm]{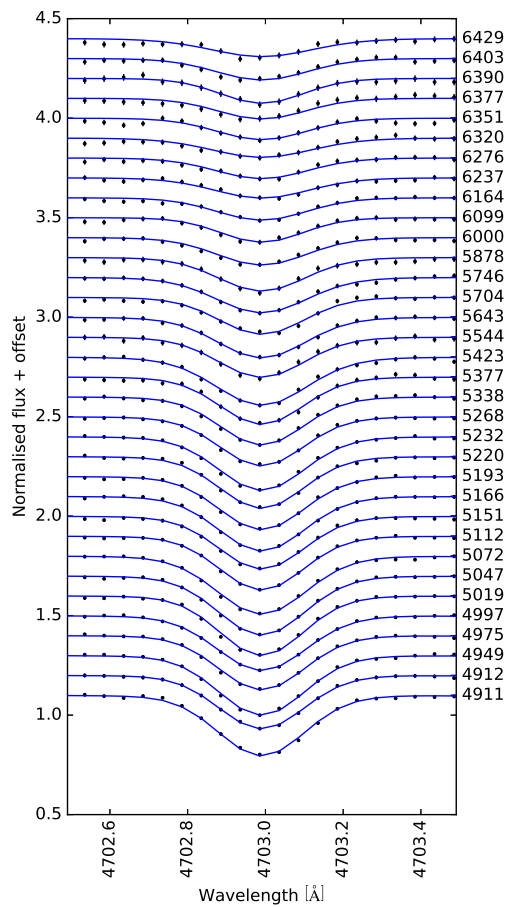}
\caption{Magnesium line for each group spectrum. Black pluses denote observed fluxes with error bars. Blue curves denote best fits given by SME assuming the $\vmic$-values given by Eqns.~\eqref{eq_vmic_dwarf} and~\eqref{eq_vmic_giant}. The vertical scale is arbitrary. On the right side are shown approximate $\teff$-values of the group spectra.}
\label{ill_Mg_fits}
\end{figure}

There is an apparent rise in the abundances near the TOP. This coincides with an increase in the width of the systematic uncertainty. We show the observed and modelled flux for the Mg line for each group spectrum in Fig.~\ref{ill_Mg_fits}. This shows the likely reason why: The line becomes much shallower near the TOP, which leads to a systematic error separate from the continuum fitting or $\vmic$, as described in Sect.~\ref{sec_systerror_source}.

\begin{table}
\caption{$\daicc$ values for Mg4702 line, assuming 1D-NLTE. Values above $10$ are written in grey, since they represent models that can essentially be ruled out.}\label{table_line_Mg4702_spectype_Combined_LOGG_NLTE_mean5500_new_vmic_constraining_systematics_error_floor_0.100}
\centering
\begin{tabular}{ l | c | c | c | c | c | c | c }
$T_0$: & 5.8 & 5.9 & 5.95 & 6.0 & 6.09 & 6.2 & null \\
\hline 
const. & $\textcolor{gray}{121}$ & $\textcolor{gray}{56.3}$ & $\textcolor{gray}{37.2}$ & $\textcolor{gray}{22.1}$ & $6.84$ & $0.00$ & $6.00$ \\
\hline 
lin. & $\textcolor{gray}{111}$ & $\textcolor{gray}{47.5}$ & $\textcolor{gray}{28.4}$ & $\textcolor{gray}{13.5}$ & $2.19$ & $1.55$ & $2.10$ \\
\hline 
quad. & $\textcolor{gray}{110}$ & $\textcolor{gray}{47.9}$ & $\textcolor{gray}{29.8}$ & $\textcolor{gray}{15.8}$ & $4.66$ & $3.96$ & $3.60$ \\
\hline 
cubic & $\textcolor{gray}{111}$ & $\textcolor{gray}{48.7}$ & $\textcolor{gray}{30.4}$ & $\textcolor{gray}{16.5}$ & $6.80$ & $6.71$ & $6.36$ \\
\hline 
4th & $\textcolor{gray}{113}$ & $\textcolor{gray}{51.7}$ & $\textcolor{gray}{33.4}$ & $\textcolor{gray}{19.3}$ & $9.29$ & $8.16$ & $8.97$ \\
\hline 
5th & $\textcolor{gray}{116}$ & $\textcolor{gray}{54.7}$ & $\textcolor{gray}{36.0}$ & $\textcolor{gray}{21.6}$ & $9.94$ & $8.06$ & $\textcolor{gray}{10.6}$ \\
\end{tabular}
\end{table}

In summary, it seems that this line on its own cannot rule out the null model, but if some external constraint does, it gives most of the information that constrains $T_0$. This is a bit surprising, given that it is the element that is expected to have the largest intrinsic scatter due to anticorrelations. As we will see in Sect.~\ref{sec_res_conc}, the reason is mostly that it has a more distinct predicted trend than the other lines, and that it is less sensitive to $\vmic$.

\subsection{Titanium}\label{sec_res_Ti}
\begin{figure}
\centering
\includegraphics[width=8.8cm]{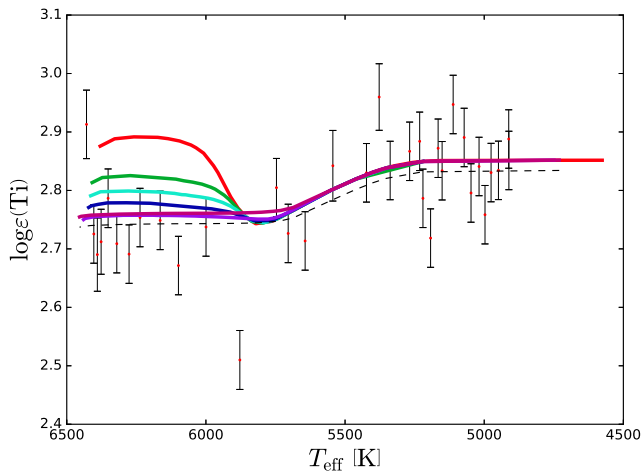}
\caption{Predictions and estimated abundances for the Ti4563 line. Symbols are the same as in Fig.~\ref{ill_Mg4702_Full_model_fit}.}
\label{ill_Ti4563_Full_model_fit}
\end{figure}

In Fig.~\ref{ill_Ti4563_Full_model_fit} we show the abundances as functions of $\teff$, together with the model predictions and the bounds on the systematic errors. The group spectrum with $\teff = 5878\,\unit{K}$, which covers the $\logg$ range $3.666$-$3.726\,\unit{dex}$, looks like an outlier since it has the derived abundance range $2.42$-$2.62\,\unit{dex}$ while the abundances for the surrounding spectra lie above $2.7\,\unit{dex}$. Based on visual inspection, we believe that this is due to a discontinuity in the observed fluxes at wavelengths close to the Ti4563 line in that spectrum, placing it among those most affected by the systematic error introduced by the continuum fitting (see Sect.~\ref{sec_systerror_source}). The same can be said about the hottest spectrum, at $\teff = 6429\,\unit{K}$, which covers the $\logg$ range $4.066$-$4.21\,\unit{dex}$.

In Table~\ref{table_line_Ti4563_spectype_Combined_LOGG_NLTE_mean5500_new_vmic_constraining_systematics_error_floor_0.100} we show the $\daicc$ values for the Ti4563 line. As one might expect from the figure, $\daicc$ is compatible with any stellar evolution model except T$5.8$. However, it can definitely rule out the null model, which the Mg4702 on its own could not do.

\begin{table}
\caption{$\daicc$ values for Ti4563 line. Values above $10$ are written in gray, since they represent models that can essentially be ruled out.}\label{table_line_Ti4563_spectype_Combined_LOGG_NLTE_mean5500_new_vmic_constraining_systematics_error_floor_0.100}
\centering
\begin{tabular}{ l | c | c | c | c | c | c | c }
$T_0$: & 5.8 & 5.9 & 5.95 & 6.0 & 6.09 & 6.2 & null \\
\hline 
const. & $\textcolor{gray}{38.2}$ & $\textcolor{gray}{11.2}$ & $4.28$ & $1.51$ & $0.91$ & $2.64$ & $\textcolor{gray}{32.5}$ \\
\hline 
lin. & $\textcolor{gray}{20.6}$ & $2.18$ & $0.99$ & $1.51$ & $2.77$ & $4.25$ & $\textcolor{gray}{20.8}$ \\
\hline 
quad. & $\textcolor{gray}{21.1}$ & $4.63$ & $3.44$ & $3.61$ & $4.18$ & $5.49$ & $\textcolor{gray}{14.8}$ \\
\hline 
cubic & $\textcolor{gray}{17.9}$ & $0.78$ & $0.00$ & $0.42$ & $0.95$ & $1.69$ & $\textcolor{gray}{14.5}$ \\
\hline 
4th & $\textcolor{gray}{20.7}$ & $3.23$ & $1.77$ & $2.23$ & $3.11$ & $3.89$ & $\textcolor{gray}{16.5}$ \\
\hline 
5th & $\textcolor{gray}{20.9}$ & $5.68$ & $4.33$ & $4.48$ & $5.06$ & $5.62$ & $\textcolor{gray}{19.5}$ \\
\end{tabular}
\end{table}

\subsection{Iron}\label{sec_res_Fe}
\begin{figure}
\centering
\includegraphics[width=8.8cm]{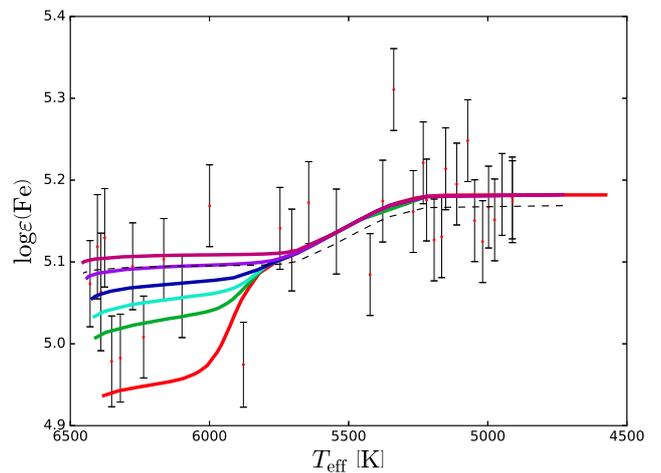}
\caption{Predictions and estimated abundances for the Fe4583 line. % Do not change the wavelength. We tend to call that line the 4583 line, but that's because we usually round down instead of to the closest integer.
Symbols are the same as in Fig.~\ref{ill_Mg4702_Full_model_fit}.}
\label{ill_Fe4583_Full_model_fit}
\end{figure}

In Fig.~\ref{ill_Fe4583_Full_model_fit} we show the abundances as a function of $\teff$, together with the model predictions and the bounds on the systematic errors. In Table~\ref{table_line_Fe4583_spectype_Combined_LOGG_NLTE_mean5500_new_vmic_constraining_systematics_error_floor_0.100} we show $\daicc$ for the fits. The line turns out to be compatible with every model, including just barely T$5.8$ and the null model. This seems to be due to a combination of large statistical errors in the derived abundances, and all models except T$5.8$ and the null model making very similar predictions.

\begin{table}
\caption{$\daicc$ values for Fe4583 line. Values above $10$ are written in grey, since they represent models that can essentially be ruled out.}\label{table_line_Fe4583_spectype_Combined_LOGG_NLTE_mean5500_new_vmic_constraining_systematics_error_floor_0.100}
\centering
\begin{tabular}{ l | c | c | c | c | c | c | c }
$T_0$: & 5.8 & 5.9 & 5.95 & 6.0 & 6.09 & 6.2 & null \\
\hline 
const. & $\textcolor{gray}{26.6}$ & $3.99$ & $0.77$ & $0.00$ & $1.55$ & $4.07$ & $\textcolor{gray}{26.2}$ \\
\hline 
lin. & $8.81$ & $3.63$ & $2.68$ & $2.40$ & $2.72$ & $3.32$ & $8.16$ \\
\hline 
quad. & $\textcolor{gray}{11.4}$ & $6.20$ & $5.20$ & $4.65$ & $4.46$ & $4.68$ & $\textcolor{gray}{10.7}$ \\
\hline 
cubic & $\textcolor{gray}{13.6}$ & $8.36$ & $7.42$ & $6.84$ & $6.59$ & $6.70$ & $\textcolor{gray}{11.5}$ \\
\hline 
4th & $\textcolor{gray}{14.8}$ & $\textcolor{gray}{11.3}$ & $\textcolor{gray}{10.3}$ & $9.56$ & $9.13$ & $9.16$ & $\textcolor{gray}{14.0}$ \\
\hline 
5th & $\textcolor{gray}{18.0}$ & $\textcolor{gray}{14.5}$ & $\textcolor{gray}{13.4}$ & $\textcolor{gray}{12.7}$ & $\textcolor{gray}{12.2}$ & $\textcolor{gray}{12.3}$ & $\textcolor{gray}{16.5}$ \\
\end{tabular}
\end{table}

\section{Summary and conclusion}\label{sec_res_conc}
We have derived stellar abundances of Mg, Ti, and Fe for a sample of stars from the metal-poor cluster M30 using a spectral fitting method. We have used the derived abundances to fit combinations of stellar evolution models and models of our systematic errors. The stellar evolution models predict various degrees of surface depletion of elements due to atomic diffusion, leading to a temperature dependence of the abundances in the visible stellar atmosphere. The models differ in the assumed efficiency of the AddMix, which is parametrised by a reference temperature $T_0$. The models of our systematic errors assume that our derived abundances are offset by a low-order polynomial in the effective temperature, and that this offset comes mostly from errors in the continuum placement and assumed microturbulence. We have then compared the combined fits using the Akaike information criterion and find that the model combinations labelled as most likely to be closest to reality are all in a small range of $T_0$.

\begin{figure}
\centering
\includegraphics[width=8.8cm]{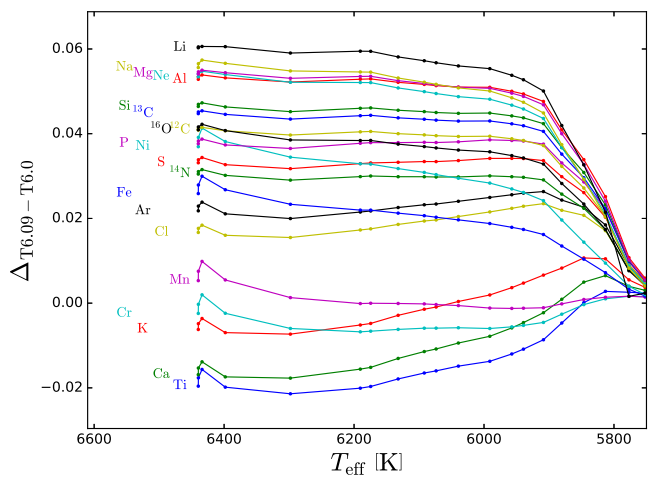}
\caption{Sensitivity of the predicted surface abundances to the choice of~$T_0$, given here as $\Delta_{\mathrm{T}6.09 - \mathrm{T}6.0}$ for each element that we have models for.}
\label{ill_trend_comparison}
\end{figure}

The process of developing our statistical method allows us to state some general conclusions about this type of study. There are three properties that a spectral line should have to be useful in constraining~$T_0$ (or some other parameter that stellar evolution models are sensitive to). First of all, the predicted trend should be as sensitive to $T_0$ as possible -- that is, $\partial \abtrend / \partial T_0$ should be as large as possible. Secondly, this sensitivity should change as quickly with the temperature as possible -- that is, $\partial^2 \abtrend / \partial T_0 \partial \teff$ should have a large value somewhere, although it does not particularly matter where. This is necessary to allow distinguishing the measured trend from a systematic error: If the stellar evolution models predict that changing $T_0$ simply raises the trend uniformly, it is very difficult to distinguish that from a systematic error. On the other hand, if the models predict trends that sharply diverge at some specific temperature, and it turns out that the measured abundances do change at that specific temperature by the amount predicted by some value of $T_0$, then it is very unlikely that this is because the systematic errors so happen to change sharply at that specific temperature. Thirdly, the measured abundances should have as little sensitivity to $\vmic$ -- and other difficult-to-constrain parameters -- as possible. Fig.~\ref{ill_trend_comparison} shows the differences $\Delta_{\mathrm{T}6.09 - \mathrm{T}6.0} \equiv \log \varepsilon \left( X_{\mathrm{T6.09}} \right) - \log \varepsilon \left( X_{\mathrm{T6.0}} \right)$, where $X_{\mathrm{Tx.y}}$ is the abundance of element $X$ assuming AddMix model $Tx.y$. Trends are shown for each element that we have models for. In most cases, the elements that have the greatest $\Delta_{\mathrm{T}6.09 - \mathrm{T}6.0}$ are also those for which the derivative of $\Delta_{\mathrm{T}6.09 - \mathrm{T}6.0}$ with respect to $\teff$ takes the greatest value at some point, such as Li, Na, Mg, Ne, and Al. Based on this, we recommend that studies of this sort focus on light elements. In particular, the Na~$8183-94\,\unit{\AA}$ doublet and the Al $13123-13150\,\unit{\AA}$ lines are measurable in first-generation TOP stars at this metallicity, while still not becoming strongly saturated on the RGB. These elements have the additional advantage that they have well-known atom data, and their electron structure is simple enough to make 3D~NLTE calculations feasible.

%\begin{figure}
%\centering
%\includegraphics[width=8.8cm]{ill_trend_derivative_comparison.eps}
%\caption{Derivative $\partial \Delta_{\mathrm{T}6.0 - \mathrm{T}5.8} / \partial \teff$ in abundance predictions for the seven elements whose derivatives take the highest values at some point.}
%\label{ill_trend_derivative_comparison}
%\end{figure}

We find that we can rule out the null hypothesis of no abundance variation, and that our stellar evolution model with atomic diffusion and additional mixing with an intermediate strength of $\log{ \left(T_0 / \left[ \unit{K} \right] \right) } = 6.09$-$6.2$ is most consistent with the derived abundances. This allows us to slightly amend one important result of Paper VI: The authors found an average lithium abundance on the Spite-plateau of $2.21 \pm 0.12\,\unit{dex}$, corresponding to an initial abundance of $A(Li)_\mathrm{init} = 2.48 \pm 0.10\,\unit{dex}$, which is considerably below the value of $2.72\,\unit{dex}$ predicted by current models of Big Bang nucleosynthesis~\citep{bbn}. However, they did so provisionally, based on a T$6.0$ model. For the T$6.09$-$6.2$ models, it would correspond to an initial abundance of $2.42$-$2.46\,\unit{dex}$. Surprisingly, the gap to the BBN lithium abundance is larger than the ones found in other globular clusters studied in this series of papers.

Even though a stellar solution to the cosmological lithium problem remains possible and even likely (see \citet{melendez_2010} and \citet{galah_lithium} for independent pieces of evidence), the physics of stellar internal mixing is too poorly understood to achieve a fully satisfactory agreement with precision-cosmology BBN predictions at present. The physical effects that give rise to additional mixing below the outer convection zone in Population II stars need to be modelled in the framework of hydrodynamics.

\begin{acknowledgements}
AG, UH and AJK acknowledge support from the Swedish National Space Agency and the Knut and Alice Wallenberg Foundation.

Parts of this research were conducted by the Australian Research Council Centre of Excellence for All Sky Astrophysics in 3 Dimensions (ASTRO 3D), through project number CE170100013.

PG thanks the European Science Foundation (ESF) for support in the framework of EuroGENESIS.

OR acknowledges the ``Programme National de Physique Stellaire'' (PNPS) of the CNRS/INSU co-funded by the CEA and the CNES, France, for financial support.

This work is in part based upon work from the ``ChETEC'' COST Action (CA16117), supported by COST (European Cooperation in Science and Technology).

Based on observations made with ESO Telescopes at the La Silla Paranal Observatory under programme ID 099.D-0748(A).
\end{acknowledgements}

%\newpage
\bibliographystyle{aa}
\bibliography{main}

\newpage
\begin{appendix}

\section{Spectra of individual stars}\label{app_spectra}
All spectra were downloaded from the ESO archive~\citep{eso_archive}. At this stage, the stellar spectra had been cleaned of cosmics and heliocentric correction had been applied, but the sky spectrum had not been removed~\citep{giraffe_manual}. For each observing night we also had 19-20 spectra observed with sky fibres. We averaged these together and subtracted this estimate of the sky flux from the stellar spectra observed on that night. After that, we coadded the individual exposures for each star. This was implemented in a Python module that itself makes use of the modules SciPy and Astropy~\citep{scipy, astropy_1, astropy_2}. Once the coaddition is finished, the radial velocity was estimated by comparison to the star HD~140283, using an algorithm by Ansgar Wehrhahn: First a cross-correlation is used to derive a first guess. Then a $\chi^2$-minimisation is performed, on the assumption that the spectra are now close enough for this not to get stuck in a local minimum~\citep{ansgar_github}.

A full list of the observed stars is shown in Table~\ref{table_individual_SP}, together with the number of spectra observed for each star, the estimated $\snr$ for the coadded spectrum, and the stellar parameters estimated as described in Sect.~\ref{sec_SP}.

\section{Group spectra}\label{app_combined}
For reasons explained in Sect.~\ref{sec_model_spec}, we coadded spectra of stars with similar stellar parameters to produce group spectra with roughly equal $\snr$. For each one of these, we took the best stellar parameter choice to be the mean of the stellar parameters of the constituent spectra. We show the properties of the group spectra in Table~\ref{table_big_spectable}, together with the derived abundances. The $\snr$ of those single-star spectra sets the minimum $\snr$ that all other group spectra are intended to match.

\section{Impact of assumed $\teff$}\label{sec_teff}
In the main body of this article, we assumed that the continuum normalisation and assumed $\vmic$ dominate over all other sources of systematic error. As a test of this assumption, we examined the impact of the assumed $\teff$ on the derived abundances. For three representative group spectra -- at the bottom, middle, and top of the temperature range -- we observed the effect of increasing the temperature by $100\,\unit{K}$. We dub the group spectra `TOP', `SGB', and `RGB', respectively. As the TOP spectrum, we selected the spectrum covering the $\logg$ range $3.95$-$3.972$, which at $6320\,\unit{K}$ is still cold enough that it can be shifted upwards by a $100\,\unit{K}$ without falling off the isochrone entirely.

In Table~\ref{table_teff_impact} we show the shift in stellar parameters and the resulting abundances for the three spectra when applying a shift of $100\,\unit{K}$ to the effective temperature. Compared to the full range covered by shifting the continuum level and assumed $\vmic$, as shown in Table~\ref{table_big_spectable}, we see that this source of error is indeed small by comparison.

%\begin{table}[h]
%\centering
%\begin{tabular}{ l | c | c | c  }
%Group spec. & RGB & SGB & TOP \\ 
%\hline
%$\teff\,\left[ \unit{K} \right]$ & $4911\to5011$ &$5643\to5743$ & $6320\to6420$ \\
%$\logg\,\left[ \unit{dex} \right]$ & $2.01\to2.29$ & $3.56\to3.62$ & $3.96\to4.13$ \\
%$\vmic\,\left[ \unit{km/s} \right]$ & $1.80\to1.71$ & $1.27\to1.30$ & $1.48\to1.46$ \\
%$\log \varepsilon{\left( \text{Fe4583} \right)}\,\left[ \unit{dex} \right]$ & $5.18\to5.30$ & $5.17\to5.19$ & $4.98\to5.06$\\
%$\log \varepsilon{\left(\text{Mg4702} \right)}\,\left[ \unit{dex} \right]$ & $5.29\to5.34$ & $5.43\to5.50$ & $5.34\to5.38$ \\
%$\log \varepsilon{\left(\text{Ti4563} \right)}\,\left[ \unit{dex} \right]$ & $2.89\to3.07$ & $2.71\to2.75$ & $2.71\to2.80$ \\
%%$\log \varepsilon{\left(\text{Ti4571} \right)}\,\left[ \unit{dex} \right]$ & $3.25\to3.44$ & $2.90\to2.92$ & $2.79\to2.88$ \\
%\end{tabular}
%\caption{The change in stellar parameters and line abundances given by shifting the temperature upwards by $100\,\unit{K}$, for three representative group spectra.}\label{table_teff_impact}
%\end{table}

\begin{table}[h]
\caption{Change in stellar parameters and line abundances given by shifting the temperature upwards by $100\,\unit{K}$, for three representative group spectra.}\label{table_teff_impact}
\centering
\begin{tabular}{ l | c | c | c }
Group spec. & RGB & SGB & TOP \\ 
\hline
$\Delta \teff\,\left[ \unit{K} \right]$ & $+100$ & $+100$ & $+100$ \\
$\Delta \logg\,\left[ \unit{dex} \right]$ & $+0.28$ & $+0.06$ & $+0.17$ \\
$\Delta \vmic\,\left[ \unit{km/s} \right]$ & $-0.09$ & $+0.03$ & $-0.02$ \\
$\Delta \log \varepsilon{\left( \text{Fe4583} \right)}\,\left[ \unit{dex} \right]$ & $+0.12$ & $+0.02$ & $+0.08$\\
$\Delta \log \varepsilon{\left(\text{Mg4702} \right)}\,\left[ \unit{dex} \right]$ & $+0.05$ & $+0.07$ & $+0.04$ \\
$\Delta \log \varepsilon{\left(\text{Ti4563} \right)}\,\left[ \unit{dex} \right]$ & $+0.19$ & $+0.04$ & $+0.09$ \\
%$\log \varepsilon{\left(\text{Ti4571} \right)}\,\left[ \unit{dex} \right]$ & $3.25\to3.44$ & $2.90\to2.92$ & $2.79\to2.88$ \\
\end{tabular}
\end{table}

% The language editor objects, but in this case "there is any assumed teff" is important
As an additional check, we tested if there was any plausible $\teff$ that would cause the RGB spectrum to assume the same abundance as the TOP spectrum. We used the same RGB spectrum as before, but for the TOP spectrum we took that covering the $\logg$ range $4.038$-$4.066\,\unit{dex}$. We derived the line abundances for the TOP spectrum and then fitted $\teff$ for the RGB spectrum while keeping the abundance at this value. Throughout we kept the broadening parameter fixed at a value corresponding to $\vmac = 8\,\unit{km/s}$ and resolution $\lambda / \Delta \lambda = 23\,000$. We found that keeping the broadening parameter free caused it to be driven up to implausibly high values -- corresponding to a $\vmac$ in excess of $30\,\unit{km/s}$ -- when doing the $\teff$ fitting. We show the TOP abundance and resulting RGB $\teff$ in Table~\ref{table_teff_shift}. Since the necessary shifts are both very large and inconsistent, we conclude that the difference in abundance cannot be purely an artefact of an offset in the $\teff$ scale.

\begin{table}[h]
\caption{Derived abundance for the TOP spectrum covering the $\logg$ range $4.038$-$4.066\,\unit{dex}$, and the resulting $\teff$ in the RGB spectrum covering the $\logg$ range $2.007$-$2.083\,\unit{dex}$ when requiring that line to have the same abundance.}\label{table_teff_shift}
\centering
\begin{tabular}{ l | c | c | c  }
Line & Fe4583 & Mg4702 & Ti4563 \\ %& Ti4571 \\ 
\hline
TOP $\log \varepsilon \left( x \right)\,\left[\unit{dex} \right]$ & $5.13$ & $5.31$ & $2.86$ \\% & $3.04$ \\
Corresponding RGB $\teff\,\left[\unit{K}\right]$ & 4643 & 5668 & 4447 \\%& 4189 \\
\end{tabular}
\end{table}

\section{Comparison to 3D~LTE}\label{app_3D_comp}
\begin{figure}
\centering
\includegraphics[width=8.8cm]{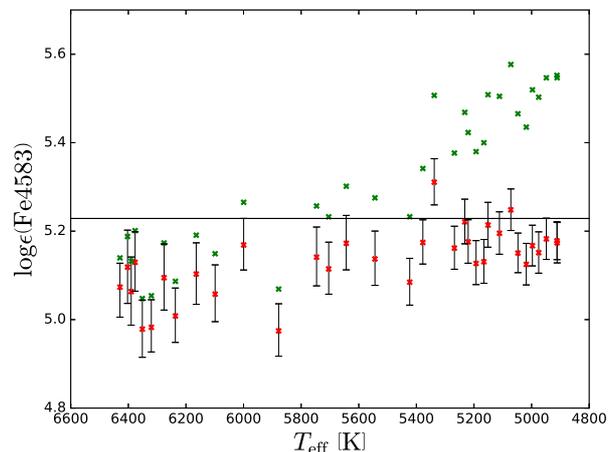}
\caption{Derived abundances from the Fe4583 line. Red crosses show 1D~LTE results and green crosses show 3D~LTE results. Error bars denote the possible systematic error due to uncertainty in the continuum fitting. The horizontal line denotes the initial abundance assumed in the stellar evolution model.}
\label{ill_Fe4583_1D_3D_comp}
\end{figure}

\begin{figure}
\centering
\includegraphics[width=8.8cm]{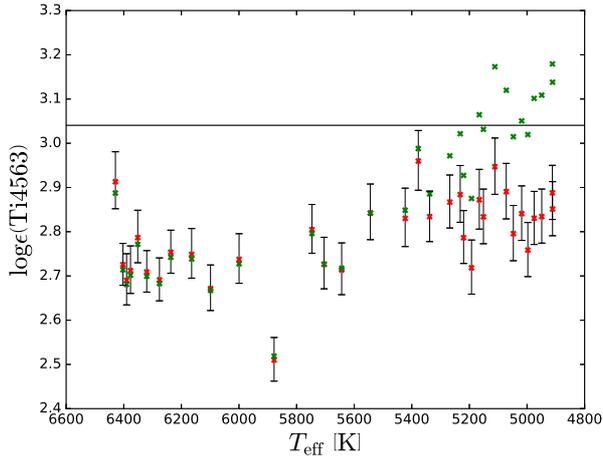}
\caption{Derived abundances from the Ti4563 line. Red crosses show 1D~LTE results and green crosses show 3D~LTE results. Error bars denote the possible systematic error due to uncertainty in the continuum fitting. The horizontal line denotes the initial abundance assumed in the stellar evolution model.}
\label{ill_Ti4563_1D_3D_comp}
\end{figure}

To test the robustness of our results, we repeated our analysis using the grids of 3D corrections for Fe and Ti described in Sect.~\ref{sec_nlte}. We find that the grids predict slightly lowered abundances at the warm end and strongly raised at the cold end. We show the change for the Fe4583 line in Fig.~\ref{ill_Fe4583_1D_3D_comp} and for the Ti4563 line in Fig.~\ref{ill_Ti4563_1D_3D_comp}.

As can be seen in the plots, the abundance trends do not appear to level out at low temperatures, which all stellar evolution models predict that they should. Since the stellar evolution models make identical predictions at the cold end, these clearly unphysical abundances do not actually affect the results of the analysis, but they make us reluctant to trust the 3D-corrected abundances for the warmer stars. We show the resulting $\daicc$ values in Table~\ref{table_bayes_spectype_Combined_LOGG_mean5500_Optimal_mix_vmic_sit_mash-sit_mash_higher-sit_mash_lower_drop_Ti4571_constraining_systematics}. The results are mostly consistent with the physical part of Table~\ref{table_bayes_spectype_Combined_LOGG_mean5500_NLTE_new_vmic_constraining_systematics_error_floor_0.100}, in that it clearly prefers T6.09 over all other models. However, it needs more parameters to model the systematic error accurately.

\begin{table}[h]
\caption{Akaike weight $w_i$, as defined by \eqref{eq_akaike_weight}, for all lines combined, after applying 3D corrections where applicable. Weights below $0.01$ are written in grey, since they represent models that, under the Bayesian interpretation of Akaike weights, have a probability below $1\%$ of being KL-minimising.}\label{table_bayes_spectype_Combined_LOGG_mean5500_Optimal_mix_vmic_sit_mash-sit_mash_higher-sit_mash_lower_drop_Ti4571_constraining_systematics}
\centering
\begin{tabular}{ l | c | c | c | c | c | c | c }
$T_0$: & 5.8 & 5.9 & 5.95 & 6.0 & 6.09 & 6.2 & null \\
\hline 
constant & $\textcolor{gray}{0.00}$ & $\textcolor{gray}{0.00}$ & $\textcolor{gray}{0.00}$ & $\textcolor{gray}{0.00}$ & $\textcolor{gray}{0.00}$ & $\textcolor{gray}{0.00}$ & $\textcolor{gray}{0.00}$ \\
\hline 
linear & $\textcolor{gray}{0.00}$ & $\textcolor{gray}{0.00}$ & $\textcolor{gray}{0.00}$ & $\textcolor{gray}{0.00}$ & $\textcolor{gray}{0.00}$ & $\textcolor{gray}{0.00}$ & $\textcolor{gray}{0.00}$ \\
\hline 
quadratic & $\textcolor{gray}{0.00}$ & $\textcolor{gray}{0.00}$ & $\textcolor{gray}{0.00}$ & $\textcolor{gray}{0.00}$ & $0.02$ & $\textcolor{gray}{0.00}$ & $\textcolor{gray}{0.00}$ \\
\hline 
cubic & $\textcolor{gray}{0.00}$ & $\textcolor{gray}{0.00}$ & $\textcolor{gray}{0.00}$ & $0.04$ & $0.83$ & $\textcolor{gray}{0.00}$ & $\textcolor{gray}{0.00}$ \\
\hline 
4th & $\textcolor{gray}{0.00}$ & $\textcolor{gray}{0.00}$ & $\textcolor{gray}{0.00}$ & $\textcolor{gray}{0.00}$ & $0.10$ & $\textcolor{gray}{0.00}$ & $\textcolor{gray}{0.00}$ \\
\hline 
5th & $\textcolor{gray}{0.00}$ & $\textcolor{gray}{0.00}$ & $\textcolor{gray}{0.00}$ & $\textcolor{gray}{0.00}$ & $\textcolor{gray}{0.00}$ & $\textcolor{gray}{0.00}$ & $\textcolor{gray}{0.00}$ \\
\end{tabular}
\end{table}

\section{Lines left out of the analysis}\label{app_unused}
Aside from the three lines discussed in the main body of the article, we fitted the lines Ba4554 and Ti4571, described in Table~\ref{table_lines_sme}. The Ba4554 line was left out from the start for the simple reason that we do not have a stellar evolution model for this element, so it cannot be used to estimate $T_0$. We only analysed it for completeness, since it was a clearly resolved line in the wavelength range we observed. We originally intended to include the Ti4571 line, but discovered that for the giant stars, it becomes heavily saturated and loses almost all its sensitivity to the abundance.

\begin{figure}
\centering
\includegraphics[width=8.8cm]{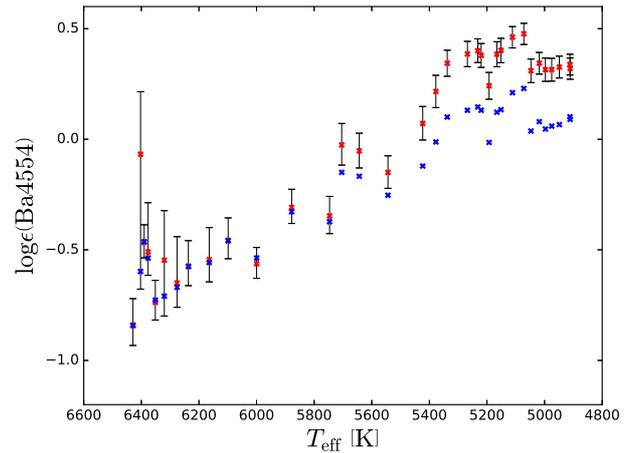}
\caption{Derived abundances for barium, based on the Ba4554 line. Red crosses show 1D~LTE and blue crosses show 1D~NLTE. Error bars denote the possible systematic error due to uncertainty in the continuum fitting.}
\label{ill_Ba_ab}
\end{figure}

In Fig.~\ref{ill_Ba_ab}, we show the derived abundances for Ba4554 assuming 1D~LTE and 1D~NLTE, with systematic error bars given by the highest and lowest possible choice of continuum. Based on this, we can see that there is a clear trend and that it levels out at low temperatures, as one would expect. The overall magnitude of the observed trend is smaller in NLTE, however.

\begin{figure}
\centering
\includegraphics[width=8.8cm]{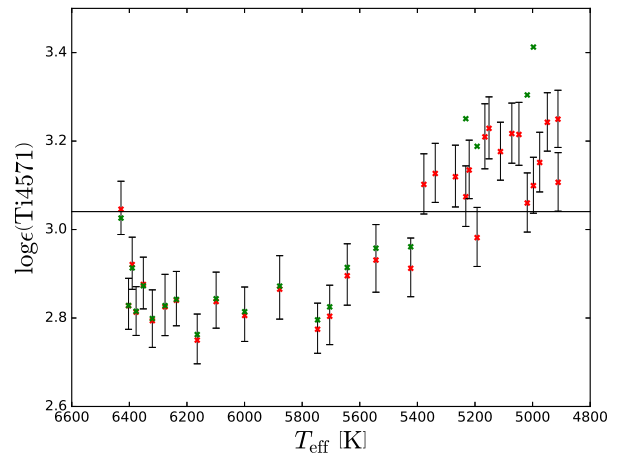}
\caption{Derived abundances for titanium, based on the Ti4571 line. Red crosses denote 1D~LTE and green crosses denote 3D~LTE. Green crosses are missing for those stars for which the parameters and derived 1D~LTE abundance fell outside of the convex hull of the grid. Error bars denote the possible systematic error due to uncertainty in the continuum fitting. The horizontal line denotes the assumed initial abundance in the stellar evolution model.}
\label{ill_Ti4571_ab}
\end{figure}

In Fig.~\ref{ill_Ti4571_ab}, we show the derived abundances for Ti4571 assuming 1D~LTE and 3D~LTE, with systematic error bars given by the highest and lowest possible choice of continuum. At low temperatures the trend does not flatten out, but rises sharply in an unphysical way. In some cases the stars fall outside the convex hull of the grid of 3D~corrections, and have to be left out completely.

\section{Impact of continuum fitting algorithm}\label{app_cont}
Current versions of SME offer two ways of estimating the continuum level: One that minimises the $\chi^2$-distance between a preliminary model spectrum and pixels covered by a continuum mask and one that compares relative line depths. The latter algorithm is a recent addition to SME, and is that used in the main body of the article. To get an estimate of the sensitivity of our analysis to the choice of algorithm, we compared our results to results using the old algorithm. We defined the continuum mask using the algorithm described in~\citet{me_1}. In Table~\ref{table_bayes_old_cont} we show the Akaike weights using this algorithm. Comparing to Table~\ref{table_bayes_spectype_Combined_LOGG_mean5500_NLTE_new_vmic_constraining_systematics_error_floor_0.100}, the preference is still for $\log_{10}{\left( T_0 / \left[ \unit{K} \right] \right)}$ in the range $6.09$--$6.2$. The main difference is that T$6.0$ is now ranked as just barely possible, and the systematic error can no longer be modelled as a constant offset. Based on this, we conclude that our analysis is robust to changes in the continuum fitting algorithm.

\begin{table}
\caption{Akaike weights $w_i$ for all lines combined, using the continuum fitting algorithm described in \citet{me_1}.}\label{table_bayes_old_cont}
\centering
\begin{tabular}{ l | c | c | c | c | c | c | c }
$T_0$: & 5.8 & 5.9 & 5.95 & 6.0 & 6.09 & 6.2 & null \\
\hline 
const. & $\textcolor{gray}{0.00}$ & $\textcolor{gray}{0.00}$ & $\textcolor{gray}{0.00}$ & $\textcolor{gray}{0.00}$ & $\textcolor{gray}{0.00}$ & $\textcolor{gray}{0.00}$ & $\textcolor{gray}{0.00}$ \\
\hline 
lin. & $\textcolor{gray}{0.00}$ & $\textcolor{gray}{0.00}$ & $\textcolor{gray}{0.00}$ & $0.01$ & $0.54$ & $0.19$ & $\textcolor{gray}{0.00}$ \\
\hline 
quad. & $\textcolor{gray}{0.00}$ & $\textcolor{gray}{0.00}$ & $\textcolor{gray}{0.00}$ & $\textcolor{gray}{0.00}$ & $0.13$ & $0.06$ & $\textcolor{gray}{0.00}$ \\
\hline 
cubic & $\textcolor{gray}{0.00}$ & $\textcolor{gray}{0.00}$ & $\textcolor{gray}{0.00}$ & $\textcolor{gray}{0.00}$ & $0.05$ & $0.02$ & $\textcolor{gray}{0.00}$ \\
\hline 
4th & $\textcolor{gray}{0.00}$ & $\textcolor{gray}{0.00}$ & $\textcolor{gray}{0.00}$ & $\textcolor{gray}{0.00}$ & $\textcolor{gray}{0.00}$ & $\textcolor{gray}{0.00}$ & $\textcolor{gray}{0.00}$ \\
\hline 
5th & $\textcolor{gray}{0.00}$ & $\textcolor{gray}{0.00}$ & $\textcolor{gray}{0.00}$ & $\textcolor{gray}{0.00}$ & $\textcolor{gray}{0.00}$ & $\textcolor{gray}{0.00}$ & $\textcolor{gray}{0.00}$ \\
\end{tabular}
\end{table}

\end{appendix}

\clearpage
\onecolumn

\setcounter{table}{0}
\def\thetable{B.\arabic{table}}

\begin{longtable}{ l | c | c | c | c| c | c | c | c | c }
\caption{Stellar parameters and line abundances for each group spectrum. For each line the smallest and greatest abundance found when varying the continuum level and assumed $\vmic$ is shown. Statistical errors are generally in the range $0.03$-$0.05$, much smaller than the systematic errors. This includes abundances that are never actually used in the model fitting, such as for the Ba4554 line.}\label{table_big_spectable}\\
 $\log{g}$ range & $T_{\mathrm{eff}}$ $\left[ \mathrm{K} \right]$ & $\left[ \mathrm{M} / \mathrm{H} \right]$ &n & $S/N$ & Ba4554 & Fe4583 & Mg4702 & Ti4563 & Ti4571 \\
\hline
\hline
2.005-2.007 & $4911$ & $-2.36$ & $1$ & $120$ & $0.01$-$0.55$& $5.03$-$5.33$& $5.37$-$5.60$& $2.60$-$3.13$& $2.83$-$3.40$\\
2.007-2.083 & $4912$ & $-2.42$ & $1$ & $104$ & $0.04$-$0.57$& $5.04$-$5.33$& $5.20$-$5.41$& $2.63$-$3.17$& $2.96$-$3.53$\\
2.083-2.126 & $4949$ & $-2.49$ & $2$ & $148$ & $0.02$-$0.56$& $5.04$-$5.34$& $5.36$-$5.58$& $2.59$-$3.11$& $2.96$-$3.52$\\
2.126-2.22 & $4975$ & $-2.41$ & $2$ & $154$ & $0.00$-$0.55$& $5.01$-$5.31$& $5.25$-$5.45$& $2.59$-$3.10$& $2.87$-$3.44$\\
2.22-2.283 & $4997$ & $-2.45$ & $2$ & $138$ & $0.00$-$0.55$& $5.03$-$5.32$& $5.32$-$5.53$& $2.52$-$3.04$& $2.83$-$3.39$\\
2.283-2.331 & $5019$ & $-2.44$ & $2$ & $134$ & $0.04$-$0.58$& $4.99$-$5.28$& $5.37$-$5.58$& $2.60$-$3.11$& $2.79$-$3.35$\\
2.331-2.414 & $5047$ & $-2.44$ & $2$ & $119$ & $-0.00$-$0.55$& $5.02$-$5.30$& $5.42$-$5.62$& $2.56$-$3.07$& $2.93$-$3.50$\\
2.414-2.467 & $5072$ & $-2.34$ & $3$ & $129$ & $0.20$-$0.68$& $5.11$-$5.41$& $5.44$-$5.65$& $2.65$-$3.16$& $2.94$-$3.50$\\
2.467-2.661 & $5112$ & $-2.37$ & $3$ & $124$ & $0.18$-$0.67$& $5.06$-$5.35$& $5.29$-$5.47$& $2.71$-$3.22$& $2.91$-$3.45$\\
2.661-2.676 & $5151$ & $-2.43$ & $3$ & $106$ & $0.10$-$0.63$& $5.08$-$5.37$& $5.31$-$5.48$& $2.61$-$3.09$& $2.96$-$3.50$\\
2.676-2.768 & $5166$ & $-2.40$ & $4$ & $117$ & $0.09$-$0.61$& $5.01$-$5.28$& $5.36$-$5.54$& $2.64$-$3.14$& $2.94$-$3.49$\\
2.768-2.842 & $5193$ & $-2.35$ & $5$ & $110$ & $-0.07$-$0.50$& $5.01$-$5.27$& $5.34$-$5.51$& $2.52$-$2.97$& $2.75$-$3.25$\\
2.842-2.909 & $5220$ & $-2.31$ & $3$ & $117$ & $0.09$-$0.61$& $5.06$-$5.32$& $5.31$-$5.46$& $2.59$-$3.03$& $2.89$-$3.40$\\
2.909-2.96 & $5232$ & $-2.36$ & $2$ & $114$ & $0.11$-$0.62$& $5.10$-$5.37$& $5.37$-$5.53$& $2.67$-$3.14$& $2.83$-$3.34$\\
2.96-3.153 & $5268$ & $-2.37$ & $8$ & $107$ & $0.10$-$0.61$& $5.05$-$5.30$& $5.27$-$5.41$& $2.67$-$3.11$& $2.88$-$3.38$\\
3.153-3.275 & $5338$ & $-2.41$ & $6$ & $94$ & $0.06$-$0.58$& $5.19$-$5.46$& $5.27$-$5.40$& $2.65$-$3.06$& $2.90$-$3.38$\\
3.275-3.341 & $5377$ & $-2.42$ & $5$ & $106$ & $-0.09$-$0.48$& $5.07$-$5.31$& $5.28$-$5.41$& $2.76$-$3.20$& $2.88$-$3.35$\\
3.341-3.459 & $5423$ & $-2.44$ & $9$ & $94$ & $-0.23$-$0.36$& $4.99$-$5.21$& $5.40$-$5.53$& $2.65$-$3.05$& $2.72$-$3.14$\\
3.459-3.532 & $5544$ & $-2.44$ & $4$ & $83$ & $-0.41$-$0.17$& $5.03$-$5.27$& $5.27$-$5.39$& $2.68$-$3.06$& $2.74$-$3.17$\\
3.532-3.589 & $5643$ & $-2.42$ & $4$ & $83$ & $-0.33$-$0.27$& $5.06$-$5.31$& $5.38$-$5.49$& $2.58$-$2.89$& $2.72$-$3.12$\\
3.589-3.61 & $5704$ & $-2.45$ & $4$ & $87$ & $-0.32$-$0.32$& $5.01$-$5.24$& $5.32$-$5.43$& $2.60$-$2.90$& $2.65$-$3.01$\\
3.61-3.666 & $5746$ & $-2.43$ & $6$ & $88$ & $-0.56$-$-0.04$& $5.03$-$5.28$& $5.36$-$5.48$& $2.67$-$2.99$& $2.64$-$2.96$\\
3.666-3.726 & $5878$ & $-2.42$ & $5$ & $86$ & $-0.50$-$-0.01$& $4.89$-$5.08$& $5.35$-$5.48$& $2.42$-$2.62$& $2.72$-$3.07$\\
3.726-3.802 & $6000$ & $-2.40$ & $5$ & $86$ & $-0.68$-$-0.37$& $5.07$-$5.29$& $5.30$-$5.42$& $2.63$-$2.88$& $2.68$-$2.97$\\
3.802-3.832 & $6099$ & $-2.42$ & $4$ & $78$ & $-0.59$-$-0.23$& $4.97$-$5.17$& $5.22$-$5.34$& $2.58$-$2.79$& $2.72$-$3.00$\\
3.832-3.883 & $6164$ & $-2.39$ & $8$ & $91$ & $-0.68$-$-0.30$& $5.01$-$5.22$& $5.24$-$5.38$& $2.65$-$2.88$& $2.65$-$2.89$\\
3.883-3.917 & $6237$ & $-2.51$ & $8$ & $85$ & $-0.69$-$-0.39$& $4.93$-$5.11$& $5.23$-$5.38$& $2.67$-$2.86$& $2.73$-$2.99$\\
3.917-3.95 & $6276$ & $-2.48$ & $8$ & $84$ & $-0.78$-$-0.37$& $5.00$-$5.21$& $5.24$-$5.39$& $2.61$-$2.79$& $2.71$-$2.98$\\
3.95-3.972 & $6320$ & $-2.50$ & $6$ & $77$ & $-0.82$-$-0.24$& $4.90$-$5.08$& $5.27$-$5.42$& $2.63$-$2.80$& $2.69$-$2.93$\\
3.972-4.003 & $6351$ & $-2.48$ & $8$ & $82$ & $-0.83$-$-0.60$& $4.90$-$5.07$& $5.28$-$5.45$& $2.70$-$2.91$& $2.77$-$3.02$\\
4.003-4.025 & $6377$ & $-2.50$ & $7$ & $79$ & $-0.64$-$-0.21$& $5.04$-$5.24$& $5.26$-$5.40$& $2.63$-$2.81$& $2.72$-$2.94$\\
4.025-4.038 & $6390$ & $-2.49$ & $6$ & $72$ & $-0.57$-$-0.33$& $4.97$-$5.18$& $5.35$-$5.47$& $2.61$-$2.79$& $2.82$-$3.06$\\
4.038-4.066 & $6403$ & $-2.50$ & $7$ & $71$ & $-0.70$-$0.41$& $5.01$-$5.24$& $5.32$-$5.45$& $2.65$-$2.82$& $2.73$-$2.96$\\
4.066-4.21 & $6429$ & $-2.50$ & $8$ & $72$ & $-0.94$-$-0.70$& $4.99$-$5.16$& $5.26$-$5.48$& $2.81$-$3.04$& $2.93$-$3.20$\\

\end{longtable}

\begin{longtable}{ l | c | c | c | c | l | c}
\caption{Individual stars, the number $n$ of spectra coadded for each star, and the best estimates of $\snr$ and stellar parameters.% Spectra with $\teff < 5200\,\unit{K}$ were not used in the analysis.
}\label{table_individual_SP}\\
\hline
Star no. & n & $\snr$ & $\teff$ $\left[ \unit{K} \right]$ & $\logg$ & $\mh$ & Comment \\
\hline
18 & 33 & 47 & 5372 & 3.29 & -2.445 & \\
21 & 28 & 44 & 5562 & 3.51 & -2.4 & \\
36 & 33 & 36 & 6380 & 4.02 & -2.435 & \\
44 & 28 & 37 & 5940 & 3.73 & -2.4 & \\
101 & 33 & 37 & 6390 & 4.03 & -2.44 & \\
108 & 33 & 39 & 6263 & 3.92 & -2.505 & \\
182 & 28 & 48 & 5741 & 3.62 & -2.48 & \\
226 & 33 & 57 & 5472 & 3.43 & -2.36 & \\
227 &  5 & 45 & 5194 & 2.79 & -2.315 & \\
234 &  5 & 43 & 5192 & 2.79 & -2.355 & \\
248 & 28 & 43 & 6240 & 3.90 & -2.525 & \\
288 & 28 & 37 & 6307 & 3.95 & -2.435 & Removed from analysis \\
305 & 28 & 37 & 6375 & 4.01 & -2.5 & Removed from analysis \\
384 & 28 & 57 & 5404 & 3.35 & -2.47 & \\
385 & 33 & 45 & 6367 & 4.00 & -2.49 & \\
392 &  5 & 28 & 5357 & 3.26 & -2.28 & \\
429 & 33 & 52 & 5859 & 3.68 & -2.365 & \\
445 & 28 & 49 & 5826 & 3.67 & -2.38 & \\
450 & 28 & 40 & 6348 & 3.99 & -2.5 & Removed from analysis \\
455 & 28 & 41 & 6221 & 3.89 & -2.515 & \\
518 &  5 & 16 & 6408 & 4.05 & -2.51 & Removed from analysis \\
669 &  5 & 29 & 5353 & 3.25 & -2.45 & \\
679 & 28 & 39 & 6344 & 3.98 & -2.505 & \\
682 & 28 & 42 & 6333 & 3.97 & -2.43 & \\
751 &  5 & 67 & 5065 & 2.41 & -2.445 & \\
848 & 33 & 50 & 6139 & 3.84 & -2.4 & \\
864 & 15 & 35 & 5926 & 3.72 & -2.44 & \\
889 & 28 & 41 & 6286 & 3.93 & -2.4 & \\
898 & 28 & 55 & 5497 & 3.46 & -2.38 & \\
902 &  5 & 18 & 6165 & 3.85 & -2.37 & Removed from analysis \\
956 & 33 & 110 & 5232 & 2.91 & -2.315 & \\
1048 &  5 & 62 & 5114 & 2.55 & -2.41 & \\
1062 &  5 & 68 & 5079 & 2.46 & -2.295 & \\
1089 & 28 & 38 & 6397 & 4.04 & -2.54 & \\
1118 & 28 & 35 & 6436 & 4.17 & -2.52 & \\
1289 & 28 & 42 & 6190 & 3.87 & -2.3 & \\
1297 &  5 & 18 & 6232 & 3.90 & -2.49 & \\
1307 & 28 & 39 & 6416 & 4.07 & -2.55 & \\
1363 & 28 & 45 & 6065 & 3.80 & -2.425 & \\
1481 & 33 & 47 & 6242 & 3.90 & -2.485 & \\
1507 & 28 & 47 & 5678 & 3.59 & -2.405 & \\
1552 & 33 & 57 & 5716 & 3.61 & -2.52 & \\
1766 & 28 & 42 & 6257 & 3.91 & -2.515 & \\
1875 &  5 & 97 & 4958 & 2.13 & -2.485 & \\
1878 &  5 & 60 & 5072 & 2.44 & -2.275 & \\
1910 &  5 & 73 & 5083 & 2.47 & -2.4 & \\
1945 &  5 & 72 & 4956 & 2.12 & -2.575 & \\
2002 & 28 & 40 & 6397 & 4.04 & -2.51 & \\
2016 &  5 & 37 & 5268 & 3.02 & -2.45 & \\
2047 & 28 & 43 & 6364 & 4.00 & -2.49 & \\
2050 &  5 & 39 & 5260 & 3.00 & -2.31 & \\
2064 & 28 & 53 & 5766 & 3.63 & -2.46 & \\
2069 &  5 & 26 & 5405 & 3.35 & -2.48 & \\
2164 & 28 & 36 & 6387 & 4.03 & -2.525 & \\
2247 &  5 & 33 & 5307 & 3.14 & -2.515 & \\
2282 &  3 & 28 & 5273 & 3.04 & -2.335 & \\
2423 & 28 & 50 & 5941 & 3.73 & -2.38 & \\
2504 &  5 & 24 & 5452 & 3.41 & -2.45 & \\
2530 &  5 & 88 & 4993 & 2.22 & -2.49 & \\
2534 &  5 & 54 & 5156 & 2.68 & -2.395 & \\
2663 & 28 & 43 & 6131 & 3.83 & -2.405 & \\
2877 &  5 & 50 & 5158 & 2.69 & -2.425 & \\
2965 & 28 & 39 & 6351 & 3.99 & -2.575 & \\
3040 &  5 & 72 & 5060 & 2.40 & -2.33 & \\
3099 &  5 & 103 & 4911 & 2.01 & -2.36 & \\
3196 &  5 & 38 & 5248 & 2.96 & -2.255 & \\
3242 & 28 & 50 & 5708 & 3.60 & -2.42 & \\
3307 &  5 & 46 & 5192 & 2.79 & -2.4 & \\
3309 & 28 & 39 & 6380 & 4.02 & -2.54 & Removed from analysis \\
3310 & 28 & 41 & 6350 & 3.99 & -2.39 & \\
3485 & 28 & 53 & 5684 & 3.59 & -2.435 & \\
3528 &  5 & 136 & 4798 & 1.71 & -2.345 & Removed from analysis \\
3594 & 28 & 40 & 6401 & 4.04 & -2.53 & \\
3634 & 28 & 40 & 6395 & 4.04 & -2.52 & Removed from analysis \\
3711 &  5 & 54 & 5151 & 2.66 & -2.425 & \\
3776 & 28 & 41 & 6131 & 3.83 & -2.395 & \\
3872 & 28 & 34 & 6437 & 4.12 & -2.465 & \\
4007 & 28 & 49 & 6042 & 3.78 & -2.395 & \\
4021 & 28 & 36 & 6406 & 4.05 & -2.44 & \\
4338 & 28 & 36 & 6430 & 4.10 & -2.5 & \\
4383 &  5 & 38 & 5252 & 2.97 & -2.375 & \\
4385 &  5 & 80 & 5020 & 2.29 & -2.425 & \\
4509 & 28 & 45 & 5674 & 3.58 & -2.45 & \\
4577 & 28 & 63 & 5383 & 3.31 & -2.395 & \\
4622 & 28 & 44 & 6332 & 3.97 & -2.47 & \\
4630 &  5 & 21 & 6270 & 3.92 & -2.485 & Removed from analysis \\
4653 &  5 & 46 & 5210 & 2.84 & -2.265 & \\
4656 & 28 & 49 & 5919 & 3.72 & -2.48 & Removed from analysis \\
4689 &  5 & 27 & 5401 & 3.34 & -2.425 & \\
4779 &  5 & 25 & 5431 & 3.39 & -2.5 & \\
4801 &  5 & 42 & 5232 & 2.91 & -2.405 & \\
4901 &  5 & 31 & 5339 & 3.22 & -2.48 & \\
4903 &  5 & 52 & 5151 & 2.66 & -2.525 & \\
4907 & 28 & 48 & 6010 & 3.76 & -2.38 & \\
4985 &  5 & 80 & 5001 & 2.24 & -2.405 & \\
5098 & 28 & 97 & 5230 & 2.91 & -2.36 & \\
5171 &  5 & 104 & 4941 & 2.08 & -2.41 & \\
5191 & 28 & 51 & 6095 & 3.81 & -2.48 & \\
5205 &  5 & 91 & 4912 & 2.01 & -2.415 & \\
5231 & 28 & 43 & 6394 & 4.03 & -2.4 & \\
5331 & 28 & 37 & 6326 & 3.97 & -2.525 & \\
5350 & 28 & 39 & 6416 & 4.07 & -2.41 & \\
5363 & 28 & 48 & 6079 & 3.80 & -2.42 & \\
5364 & 28 & 32 & 6426 & 4.20 & -2.55 & \\
5379 &  5 & 43 & 5186 & 2.77 & -2.375 & \\
5390 & 28 & 44 & 6201 & 3.88 & -2.52 & \\
5410 &  5 & 83 & 5017 & 2.28 & -2.46 & \\
5457 & 28 & 42 & 6276 & 3.93 & -2.475 & \\
5458 & 28 & 44 & 6316 & 3.96 & -2.55 & \\
5473 &  5 & 17 & 6361 & 4.00 & -2.475 & Removed from analysis \\
5523 &  5 & 89 & 4993 & 2.22 & -2.34 & \\
5712 &  5 & 54 & 5165 & 2.71 & -2.395 & \\
5719 & 28 & 39 & 6366 & 4.00 & -2.56 & \\
5732 & 28 & 49 & 6115 & 3.82 & -2.35 & \\
5733 & 28 & 51 & 5632 & 3.56 & -2.385 & \\
5792 &  5 & 21 & 5797 & 3.65 & -2.39 & Removed from analysis \\
5804 &  5 & 51 & 5183 & 2.76 & -2.375 & \\
5885 &  5 & 57 & 5152 & 2.67 & -2.33 & \\
5950 & 28 & 47 & 6234 & 3.90 & -2.475 & \\
5976 &  5 & 21 & 6256 & 3.91 & -2.565 & \\
5980 &  5 & 28 & 5375 & 3.30 & -2.4 & \\
5987 & 28 & 41 & 6281 & 3.93 & -2.45 & \\
6006 & 15 & 26 & 6438 & 4.16 & -2.5 & Removed from analysis \\
6044 &  5 & 73 & 5035 & 2.33 & -2.545 & \\
6099 &  5 & 41 & 5255 & 2.98 & -2.325 & \\
6105 & 28 & 35 & 6433 & 4.18 & -2.48 & Removed from analysis \\
6142 &  5 & 43 & 5220 & 2.87 & -2.295 & \\
6218 & 28 & 52 & 5728 & 3.61 & -2.375 & \\
6231 &  5 & 30 & 5344 & 3.23 & -2.35 & \\
6239 & 28 & 44 & 6268 & 3.92 & -2.58 & \\
6258 &  5 & 20 & 6280 & 3.93 & -2.505 & \\
6292 &  5 & 29 & 5365 & 3.28 & -2.375 & \\
6309 & 28 & 45 & 5709 & 3.60 & -2.425 & \\
6313 & 28 & 76 & 5321 & 3.17 & -2.39 & \\
6328 & 28 & 39 & 6388 & 4.03 & -2.53 & \\
6419 &  5 & 18 & 6360 & 4.00 & -2.495 & Removed from analysis \\
6441 & 28 & 41 & 6382 & 4.02 & -2.495 & \\
6460 &  5 & 30 & 5281 & 3.06 & -2.4 & \\
6519 &  5 & 18 & 6171 & 3.86 & -2.4 & Removed from analysis \\
6614 &  2 & 17 & 5411 & 3.36 & -2.39 & Removed from analysis \\
6621 & 28 & 45 & 6186 & 3.87 & -2.3 & \\
6654 & 28 & 52 & 5590 & 3.53 & -2.44 & \\
6678 &  5 & 20 & 5721 & 3.61 & -2.43 & Removed from analysis \\
6714 &  5 & 44 & 5199 & 2.81 & -2.285 & \\
6730 &  5 & 24 & 5428 & 3.38 & -2.365 & \\
67415 &  28 & 74 & 5393 & 3.33 & -2.49 & \\
200065 & 33 & 33 & 6386 & 4.03 & -2.535 & Removed from analysis \\
200096 & 28 & 36 & 6329 & 3.97 & -2.45 & Removed from analysis \\
200249 &  5 & 17 & 6213 & 3.88 & -2.485 & Removed from analysis \\
200275 &  5 & 23 & 5404 & 3.35 & -2.495 & \\
200512 & 28 & 40 & 6312 & 3.95 & -2.545 & \\
200791 & 28 & 44 & 5725 & 3.61 & -2.425 & \\
200809 & 28 & 43 & 5861 & 3.68 & -2.43 & \\
200824 &  5 & 53 & 5137 & 2.62 & -2.31 & \\
200868 &  5 & 22 & 5313 & 3.15 & -2.495 & \\
200907 & 28 & 43 & 5582 & 3.53 & -2.515 & \\
201211 & 28 & 40 & 6106 & 3.82 & -2.425 & \\
201548 & 30 & 51 & 5534 & 3.49 & -2.46 & Removed from analysis \\
201705 & 28 & 35 & 6402 & 4.05 & -2.485 & \\
201887 & 28 & 35 & 6408 & 4.05 & -2.465 & \\
201979 & 28 & 34 & 6386 & 4.02 & -2.455 & \\
201997 & 33 & 40 & 6284 & 3.93 & -2.44 & Removed from analysis 
\end{longtable}

\end{document}